\documentclass[twocolumn,prd,superscriptaddress,nofootinbib]{revtex4-1}

\usepackage{graphicx,verbatim}
\usepackage{amsmath,bm}
\usepackage[usenames,dvipsnames]{xcolor}
\usepackage[normalem]{ulem}
\usepackage{url}
\usepackage{multirow}
\usepackage[colorlinks  = true,
            linkcolor   = NavyBlue,
            urlcolor    = NavyBlue,
            citecolor   = NavyBlue,
            anchorcolor = NavyBlue]{hyperref}

\def \aap  {A\&A}

\def \apj  {ApJ}

\def \apjl  {ApJL}

\def \prd {Phy. Rev. D}

\def \mnras {MNRAS}

\def \prl {PRL}

\begin{document}

\title{Self-confinement of relativistic pair beams in magnetized interstellar plasmas: \\the case of pulsar X-ray filaments}

\author{Luca Orusa}
\email{luca.orusa@princedon.edu} 
\affiliation{Department of Astronomy and Columbia Astrophysics Laboratory, Columbia University, New York, NY, 10027, USA}
\affiliation{Department of Astrophysical Sciences, Princeton University, Princeton, NJ, 08544, USA} 
\author{Lorenzo Sironi}
\email{lsironi@astro.columbia.edu} 
\affiliation{Department of Astronomy and Columbia Astrophysics Laboratory, Columbia University, New York, NY, 10027, USA}
\affiliation{Center for Computational Astrophysics, Flatiron Institute, 162 5th avenue, New York, NY, 10010, USA}

\begin{abstract}
The observation of filamentary X-ray structures near bow-shock pulsar wind nebulae (PWNe)---such as the Guitar, Lighthouse, and PSR J2030+4415 nebulae---and of slow-diffusion regions around pulsars like Geminga, Monogem, and PSR J0622+3749, challenges the standard picture of cosmic-ray transport in the interstellar medium, implying a diffusion coefficient two orders of magnitude smaller than the Galactic average. The suppressed diffusion can be attributed to self-generated magnetic turbulence, driven---via the non-resonant streaming instability---by electron–positron pairs escaping the PWNe. This instability requires a net current, yet the beam of escaping pairs is expected to be charge-neutral. We show that a charge-neutral pair beam propagating through an electron--proton plasma can spontaneously generate a net current. Using fully kinetic two-dimensional particle-in-cell simulations with realistic mass ratio, we find that beam electrons get focused into self-generated magnetic filaments produced by the nonlinear evolution of the Weibel instability, while beam positrons remain unconfined. We show that in three-dimensional simulations the resulting net (positron) current drives the non-resonant streaming instability, further amplifying the magnetic field. This mechanism provides a pathway for the onset of charge asymmetries in initially charge-neutral pair beams and for the growth of magnetic fluctuations that efficiently scatter the beam particles, with implications for the formation of X-ray filaments and, potentially, for particle self-confinement in TeV halos around PWNe.

\end{abstract}

\maketitle

\noindent\textbf{\textit{Introduction.---}}
Filamentary X-ray structures, also called kinetic jets or misaligned outflows---several parsecs long and less than a parsec wide---have been detected around multiple bow-shock pulsar wind nebulae (PWNe) \citep{Hui_Becker:2007,Pavan:2014,Temim:2015,Klingler:2016,Klingler:2018,Medvedev:2019,Marelli:2019,Bordas_Zhang:2020,Zhang:2020,Klingler_Yang:2020,deVries-J2030:2022,dinsmore+24}, most notably the Guitar \citep{Hui_Becker:2007,deVries:2022,dinsmore+25} and Lighthouse nebulae \citep{Pavan:2014,Pavan:2016,Klingler:2023}. In \cite{olmi+24} their morphologies and spectra are attributed to synchrotron emission from relativistic leptons streaming along magnetic fields that are stronger and more turbulent than those of the ambient interstellar medium (ISM). This requires both efficient magnetic field amplification and significant suppression of the local diffusion coefficient. A leading mechanism capable of strongly amplify the field is the non-resonant (Bell) instability (NRI) \citep{Bell:2004} (but see \cite{plotnikov+24} for a recent study focused on the resonant streaming instability), which arises from a net beam current and amplifies magnetic fields on sub-Larmor scales before inverse cascading to larger scales \citep{Bandiera:2008,olmi+24}. A key question for triggering the NRI is the origin of the net beam current, since pulsar winds produce comparable numbers of electrons and positrons ($e^\pm$). 

\citet{Olmi_Bucciantini:2019a} performed magnetohydrodynamic simulations and showed that leptons with energies close to the maximum potential drop of the pulsar can escape the nebula as a charge-separated flow. However, the escape of lower-energy particles, which are more abundant and thus dominate the current, is quasi-neutral. Therefore, it is essential to investigate whether a pair beam with zero net current can develop strong self-confinement. Filamentary X-ray structures near PWNe may thus trace regions of magnetic field amplification and inhibited diffusion. This picture parallels the TeV halos detected around middle-aged pulsars (e.g., Geminga, Monogem) by HAWC \citep{hawc17,hawc23}, Milagro \citep{milagro09}, HESS \citep{hess23}, and LHAASO \citep{lhaaso21,cao+24}, and in the GeV band by Fermi-LAT \citep{dimauro+19}. Both TeV halos and X-ray filaments exhibit diffusion coefficients suppressed by a factor of $10^2$--$10^3$ relative to the Galactic average, suggesting that magnetic turbulence driven by the escaping pulsar pairs may self-regulate their propagation \citep{olmi+24,dinsmore+26, Bandiera:2008,evoli+18,Mukhopadhyay22,plotnikov+24}, with implications for the interpretation of the positron excess in cosmic rays \citep{orusa+21,orusa+25}. Alternative explanations proposed for TeV halos include asymmetric diffusion \citep{liu+19,DeLaTorreLuque+22}, ballistic transport \citep{Recchia21}, and external turbulence \citep{Lopez-Coto+18,fang+19,bourguinat+25}. With improving observational capabilities across the electromagnetic spectrum, a complete understanding of the conditions under which relativistic pair-driven instabilities grow is becoming increasingly vital. 

In this work, we investigate pair-driven instabilities using 2D and 3D particle-in-cell (PIC) simulations, focusing on their long-term evolution in the context of PWNe. 
\citet{peterson+21,peterson+22} identified an instability---evident in other PIC studies \citep{golant+24,groselj+24,bresci+22,demidov+25}---termed cavitation, which emerges after the saturation of the Weibel instability \citep{weibel59} seeded by the streaming of $e^\pm$ pairs through a background electron--proton plasma. While their analysis focused on beams propagating into an unmagnetized plasma, we investigate the cavitation instability in a magnetized background where the magnetic energy density is roughly comparable to the beam energy density, as expected in flux tubes filled with escaping PWNe leptons.
We find that the cavitation mode can indeed develop when the beam energy density exceeds the thermal and magnetic energy densities of the background. Under these conditions, the instability leads to strong amplification of the transverse magnetic field.
In an initially charge- and current-neutral pair beam, the nonlinear evolution spontaneously produces a net current, as beam $e^-$ become magnetically trapped inside the magnetized cavities while beam $e^+$ stream outside. Using 3D simulations, we show that the resulting net positron current naturally drives the non-resonant (Bell) instability \cite{Bell:2004}.

\smallskip
\noindent\textbf{\textit{Simulation setup.---}} 
We perform 2D and 3D PIC simulations of a dilute $e^\pm$ beam streaming through a background $e^--p$ plasma. The beam propagates along a uniform magnetic field of strength $B_0$ oriented along $\hat{z}$. We use periodic boundaries in all directions. The computational domain in 2D lies in the $x$–$y$ plane, as done by \citet{peterson+22}. Our simulation domain is meant to represent a portion of the magnetic flux tube populated by the relativistic pairs escaping from PWNe. The beam and background plasmas are initialized at $t=0$ and evolved self-consistently. We also test an alternative setup in which beam particles are stochastically reset to their initial momentum distribution, to mimic the continuous pair injection from a neighboring PWN: at each timestep, particles are ``refreshed'' with a probability set by the requirement that, on average, each beam particle is re-drawn from its initial distribution over some timescale $t_{\rm refresh}$ (see End Matter). The background electron and ion densities are $n_{0,i}=n_{0,e^-}=n_0$, while the beam density is $n_b=n_{b,e^+}=n_{b,e^-}$. We employ the realistic mass ratio $m_i/m_e = 1836$.
All energy densities are normalized to the rest mass energy density of background electrons, $n_0 m_e c^2$:

\begin{itemize}%[noitemsep, topsep=0pt, parsep=0pt, partopsep=1pt]
\item Beam energy: $\alpha \gamma_b$, where $\gamma_b$ is the Lorentz factor of the beam, and $\alpha$ is the beam-to-background number density ratio, $\alpha=n_b/n_0$.

\item Background magnetic energy: ${B_0^2}/{(8 \pi n_0 m_e c^2)} = \sigma/2$, where $\sigma$ is the magnetization parameter; 

\item Background thermal energy: ${k_B T_{\rm ISM}}/{m_e c^2} = \Delta \gamma$, where $T_{\rm ISM}$ is the temperature of the background plasma and $k_B$ is the Boltzmann constant. 

\end{itemize}
We assume $\sigma/2 = \Delta \gamma$ in all simulations, consistent with the conditions expected in the ISM.
We also check the dependence on the dimensionless beam temperature ${k_B T_b}/{m_e c^2} = \Delta \gamma_b$.  
The simulations are performed with the PIC code Tristan-MP \cite{spitkovsky05}, using normalizations $\omega_{pe}^{-1} = (m_e / 4\pi n_0 e^2)^{1/2}$ for time and $d_e = c / \omega_{pe}$ for length. 2D runs employ 10 cells per $d_e$, while 3D runs have 5 cells per $d_e$. Each species of the beam or background plasma is modeled with 18 macroparticles per cell (ppc) in 2D and 9 in 3D. Different beam-to-background density ratios $\alpha$ are implemented by adjusting the weight of beam particles, while keeping the same number of beam and background macroparticles per cell. The fiducial domain spans $48 \times 48\, d_e^2$ in 2D and $256 \times 256 \times 128\, d_e^3$ in 3D. Larger 2D runs with $200 \times 200\, d_e^2$ and $1000 \times 1000 \, d_e^2$ presented in the End Matter confirm our conclusions. Further tests demonstrating numerical convergence are in the End Matter.

\begin{figure}[!t]
    \centering
\includegraphics[width=0.5\textwidth,clip=true,trim= 0 625 0 0]{"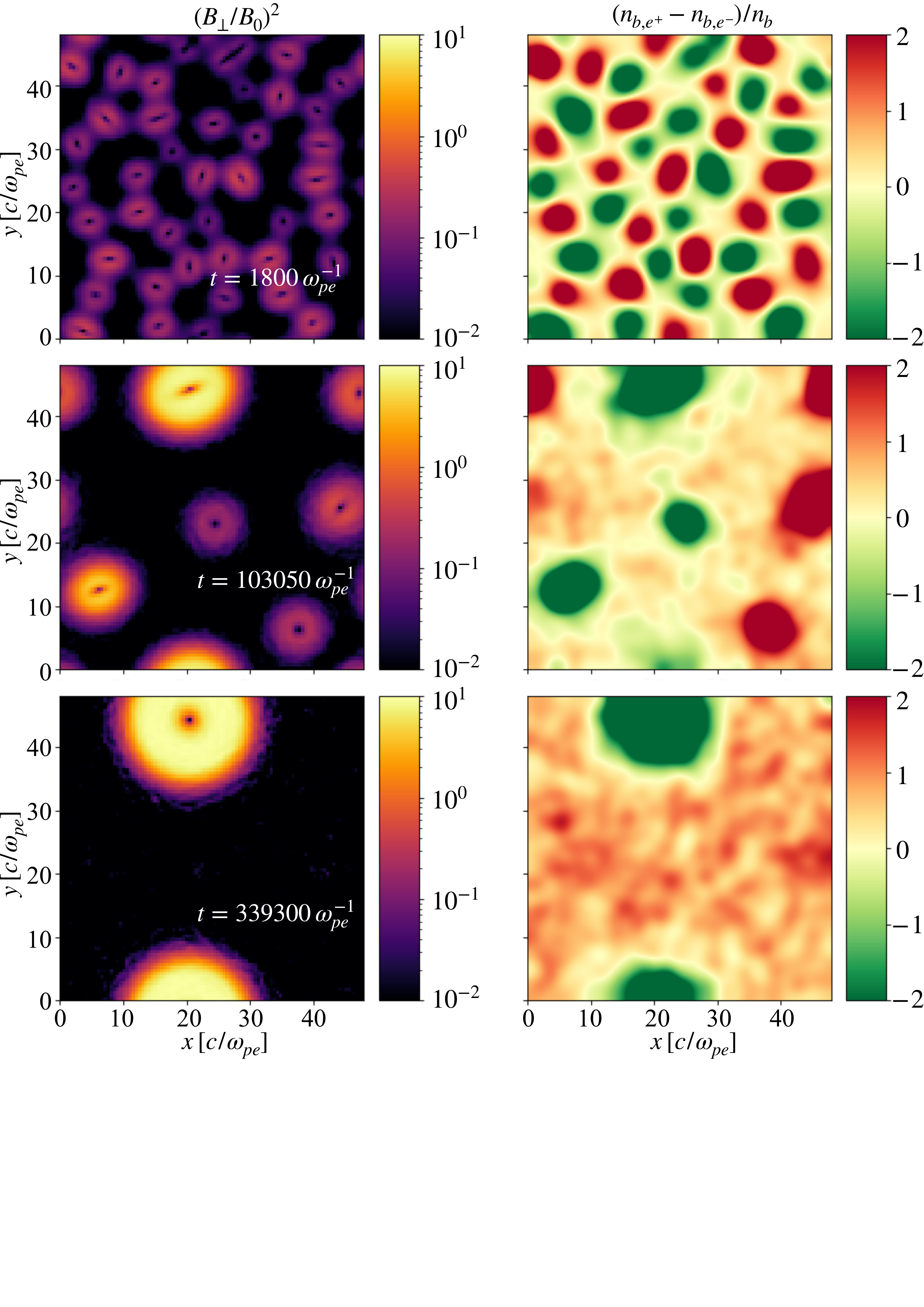"}
    \caption{Evolution of $(B_\perp/B_0)^2$ (first column) and of the charge separation (second column) of beam positrons and electrons $(n_{b,e^+}-n_{b,e^-})/n_b$, for the reference case with $\alpha \gamma_b = 10^{-1}$, $\sigma = 2 \times 10^{-2}$, and $\Delta{\gamma_b}=10^{-4}$.} 
    \label{Fig:visual}
    \vspace{-15pt}
\end{figure}

\smallskip
\noindent\textbf{\textit{Results.---}}
Figure~\ref{Fig:visual} illustrates the time evolution of the perpendicular field $(B_\perp/B_0)^2=({B_x^2+B_y^2})/B_0^2$ and the charge separation of beam $e^\pm$ for a run with $\alpha \gamma_b = 10^{-1}$ and $\sigma = 2 \times 10^{-2}$ (i.e., beam to background energy ratio of 10). The beam temperature is $\Delta\gamma_b = 10^{-4}$. Larger beam temperatures still produce consistent results, although a detailed analysis of the dependence on $\Delta\gamma_b$ is beyond the scope of this Letter.
 
The onset of the Weibel instability, driven by the streaming of the pair beam through the background electron--proton plasma, breaks the beam into current filaments composed of charge-separated $e^-$ and $e^+$. The Weibel mode saturates at $B_\perp\ll B_0$ \cite{davidson+72}. After saturation, while the current filaments of beam $e^+$  rapidly dissipate, $e^-$ filaments can only be charge-neutralized by the background ions, whose large inertia prevents efficient current screening. The magnetic pressure in the unscreened  filaments of beam $e^-$ inflates them into expanding magnetized cavities---a manifestation of the cavitation instability \cite{peterson+22}, which enhances both field strength and coherence scale by orders of magnitude.
Meanwhile, beam $e^+$ continue streaming outside the cavities, generating a volume-filling net current.

\begin{figure}[!t]
\centering
\includegraphics[width=0.49\textwidth,clip=true,trim= 0 1150 0 0]{"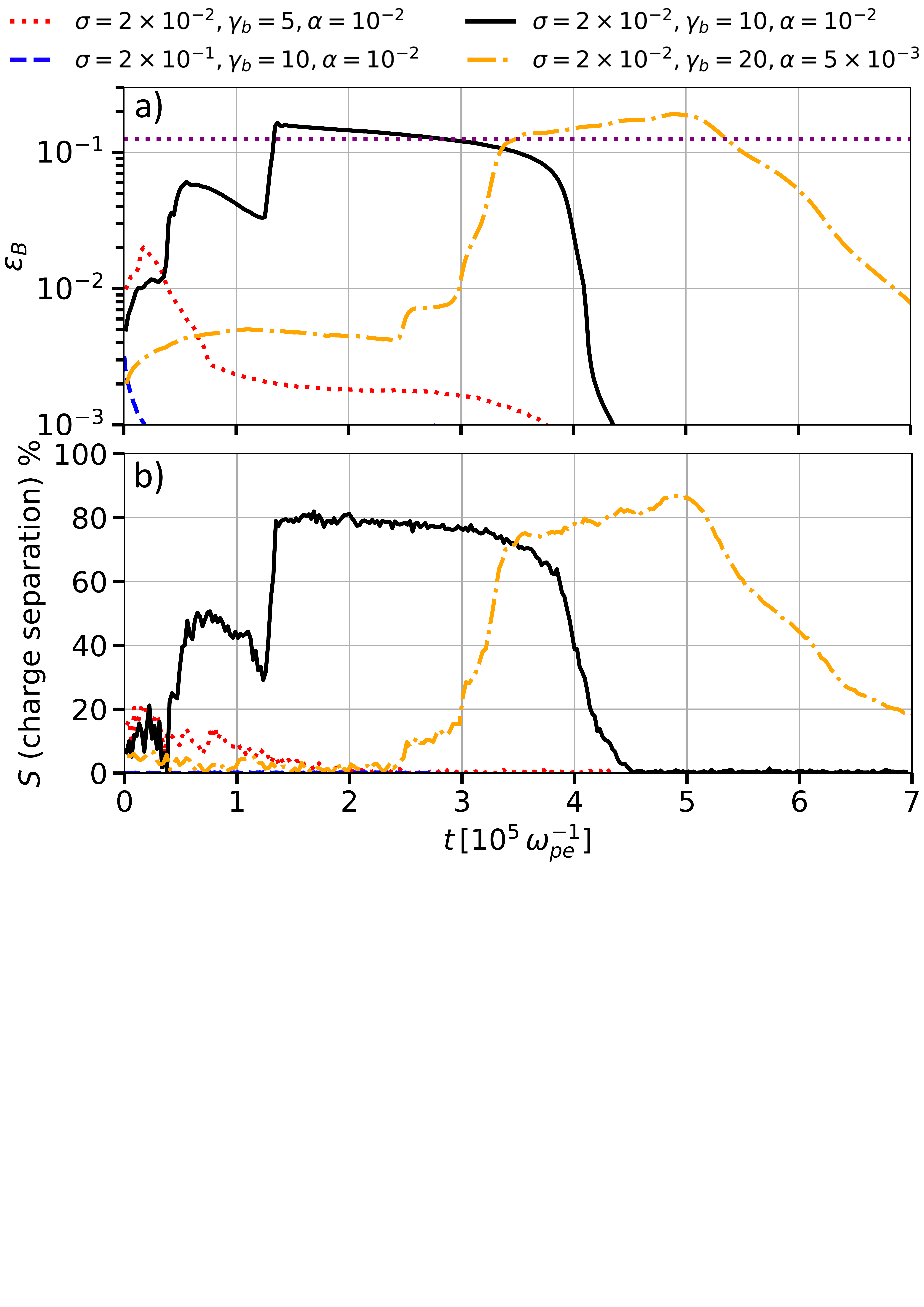"}
    \caption{Fraction of beam energy converted into magnetic energy (a) and charge separation between beam electrons and positrons (b) for various $\sigma$ and $\gamma_b\alpha$, at fixed $\Delta\gamma_b = 10^{-4}$. The horizontal purple line is the theoretical prediction in Eq. \ref{eq:saturation}.
    } 
    \label{Fig:no_refresh}
    \vspace{-15pt}
\end{figure}

Figure~\ref{Fig:no_refresh}a shows the fraction of beam energy converted into magnetic energy, $\epsilon_B=B_\perp^2/(8\pi n_0\gamma_b\alpha m_e c^2)$, for various $\sigma$ and $\gamma_b\alpha$, at fixed $\Delta\gamma_b = 10^{-4}$. Significant field amplification occurs only when the beam energy exceeds the background thermal and magnetic energy by at least an order of magnitude.
From \citet{peterson+21}, the cavitation instability saturates when
\begin{equation}\label{eq:saturation}
\epsilon_B \sim \frac{1}{8} \; {\rm min} \left\{1, \frac{m_i}{\gamma_b m_e} \right\}.
\end{equation}
which is shown by the horizontal purple dotted line in Figure~\ref{Fig:no_refresh}a.
In our reference case (black solid line), the cavitation instability saturates at $t \sim 1.5 \times 10^{5}\,\omega_{pe}^{-1}$, converting $\sim 15\%$ of the beam energy into magnetic energy, consistent with the theoretical prediction of Eq.~\ref{eq:saturation}. On the way to saturation, neighboring filaments merge (which leads to an increase in $\epsilon_B$, e.g., at $t\sim 5 \times 10^{4}\,\omega_{pe}^{-1}$), while isolated filaments decay. When a single filament remains in the box, the overall $\epsilon_B$ drops---in our reference case, on a timescale of $\sim 3 \times 10^{5}\,\omega_{pe}^{-1}$. Larger-box runs (End Matter) reach the same saturation level of $\epsilon_B$ but exhibit slower decay, suggesting longer-lived fields in larger and more realistic systems. 

The charge separation fraction outside of the cavities, defined as $S=(N_{b,e^+}-N_{b,e^-})/N_{b,tot,e^+}$, is shown in Fig.~\ref{Fig:no_refresh}b. Here, $N_{b,e^+}$ and $N_{b,e^-}$ are the total number of beam $e^+$ and $e^-$ outside of cavities (defined as regions where $(B_\perp/B_0)^2<0.04$). When the cavitation instability is triggered, $S$ can reach $\sim 80$\%.

The onset of the cavitation instability depends on the ratio of beam to background energy density, $\alpha \gamma_b / \sigma$. The growth rate of the cavitation instability scales linearly with $\alpha$ \cite{peterson+22}. Lowering $\alpha$ delays the growth and saturation of the instability, but the fraction of beam energy converted into magnetic energy remains unchanged, as does the degree of charge separation, as shown in Fig.~\ref{Fig:no_refresh} (compare black and orange). Decreasing $\alpha \gamma_b$ and $\sigma$ while maintaining their ratio constant leads to the same results for both $\epsilon_B$ and $S$ after saturation, demonstrating that $\alpha \gamma_b / \sigma$ is the key parameter governing the onset of the instability (see End Matter). For the ``refreshed'' cases shown in the End Matter, we find similar results. There, the cavitation instability develops more easily, for beam-to-background energy ratios an order of magnitude lower than in the non-refreshed setup.

From Eq.~\ref{eq:saturation}, one expects that at $\gamma_b>m_i/m_e$ the fraction of beam energy converted into magnetic energy will be lower. To test this, we employed $m_i/m_e=100$, $\alpha=10^{-2}$, and $\gamma_b=4000$ (not shown). We find that the efficiency of field amplification is indeed lower, with $\epsilon_B$ saturating at $0.0034$, close to the predicted value. Nevertheless, the degree of charge asymmetry remains strong ($S\sim83\%$ at saturation of the cavitation mode), consistent with our reference case. Thus, we conclude that whenever the cavitation instability is allowed to grow, the degree of charge separation is significant, regardless of whether $\gamma_b<m_i/m_e$ or not.

\begin{figure}[!t]
\centering
\includegraphics[width=0.49\textwidth,clip=true,trim= 0 0 150 0]{"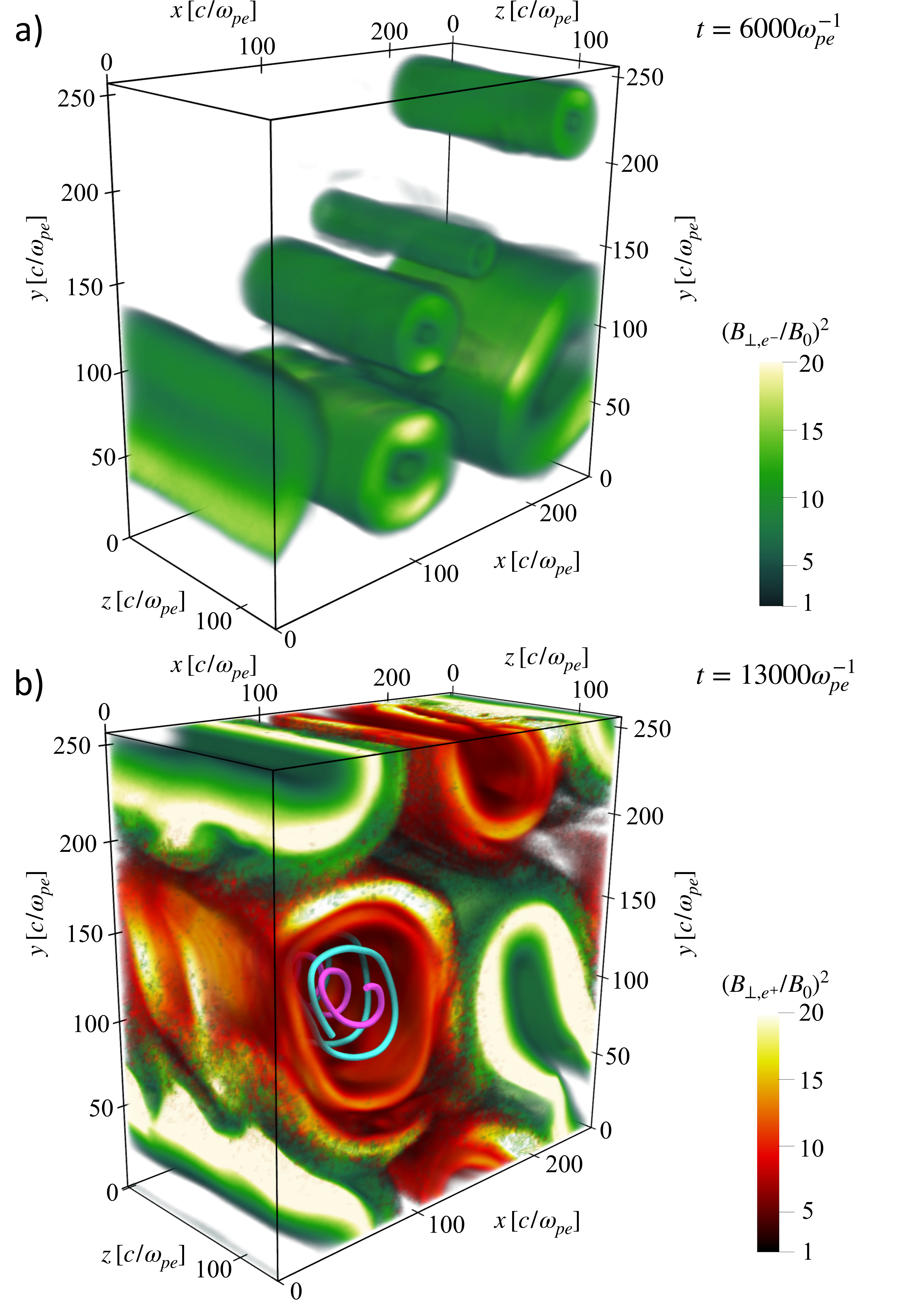"}
    \caption{(a) $(B_\perp/B_0)^2$ during the growth of the cavitation instability; (b) $(B_\perp/B_0)^2$ showing the development of the NRI (in red, driven by beam $e^+$) at later times, together with the cavities generated by beam $e^-$ (in green, as in (a)).    
    Magnetic-field streamlines within the $e^+$-driven NRI regions are also shown. We adopt $\sigma=10^{-1}$ and $\alpha \gamma_b= 5 \times 10^{-1}$.
    } 
    \label{Fig:3D}
    \vspace{-15pt}
\end{figure}

The resulting asymmetry between beam $e^+$ and $e^-$ outside of the magnetized cavities can seed current-driven instabilities along the beam direction, such as the NRI \cite{Bell:2004,gupta+21,olmi+24}. Since the NRI grows along the beam propagation direction, only fully 3D simulations can simultaneously capture the growth of the cavitation instability and the resulting development of the NRI.
Excitation of the NRI requires that the fastest-growing mode, $k_{\rm max}$, satisfies $k_{\rm max} R_L > 1$, where $R_L$ is the Larmor radius of $e^+$ in the background field $B_0$. As in \citet{olmi+24} and \citet{zacharegkas+24}, this condition can be recast as
\begin{equation}\label{eq:bell}
%\frac{S \, \alpha \gamma_b}{2} > \sigma. 
S \, \alpha \gamma_b \gg \sigma 
\end{equation}
which is satisfied in all the cases where the cavitation instability is allowed to grow ($\gamma_b\alpha\gg\sigma$, leading to $S\gtrsim 80\%$).
For a relativistic beam, the maximum growth rate of the NRI is given by \cite{gupta+21}
\begin{equation}
\frac{\Gamma_{\rm max}}{\omega_{pe}}= 
\frac{S \alpha}{2} \sqrt{\frac{m_e}{m_i}}% \sim 2 \times 10^{-4}.
\end{equation}
which is identical (apart from a factor of $S/2$) to the growth rate of the cavitation instability \cite{peterson+21}. If the cavitation instability grows, it leads to $S\sim 1$ and then the NRI will also grow on comparable timescales. The NRI growth time is much shorter than the decay time of the magnetized cavities, i.e., significant charge asymmetry is steadily sustained during the NRI growth.

To test this scenario, we perform a 3D simulation with $\gamma_b=10$, $\alpha=5 \times 10^{-2}$, $\sigma=10^{-1}$ and $\Delta \gamma_b=5$; the latter is chosen because, as shown in \cite{zacharegkas+24}, hot beams are expected to drive the NRI more efficiently than  cold beams. We set $m_i/m_e=25$ so our simulations fall in the appropriate (low-current) regime of the NRI, as expected for realistic parameters \cite{lichko+25}. The results are shown in Fig.~\ref{Fig:3D}. 
The cavitation instability grows (Fig.~\ref{Fig:3D}a), forming toroidal structures around the beam propagation direction. The transverse field is amplified to $(B_\perp / B_0)_{\rm cav} \sim 1.1$, consistent with Eq.~\ref{eq:saturation} accounting for the fact that  cavities fill half of the volume. 
At later times, beam $e^+$ streaming outside the cavities further amplify the field via the NRI, generating right-hand circularly polarized waves with wavevector aligned with the $z$ direction of the background field. In Fig.~\ref{Fig:3D}b, the transverse magnetic field $B_\perp$ produced by the streaming $e^+$ is shown in red, while the cavities generated by beam $e^-$ are shown in green, as in (a). Magnetic field lines inside the $e^+$-dominated regions display right-hand circular polarization, a hallmark of the NRI driven by beam $e^+$ \cite{gupta+21}. Given that the NRI  wavevector is along $z$, it should not grow in 2D, as we indeed confirm in the End Matter for the same parameters as our 3D run.

The NRI saturates when the fastest-growing modes---that grows on scales of order $1/k \sim 7 \,d_e$ during the linear phase, consistent with theoretical expectations and smaller than the Larmor radius of the beam particles---transfer power to larger scales, until they reach the Larmor scale as measured with the NRI-amplified field. Equivalently, the energy in the transverse field $B_\perp$ generated by the NRI is (Eq. 10 in \cite{zacharegkas+24})
%$\Delta B$. The saturation condition $k_{\max}^* R_L = 1$ leads to a field amplification 
\begin{equation}
\left(\frac{B_\perp}{B_0}\right)^2_{\rm NRI} \sim \frac{2 S \alpha\gamma_b (1+\Delta \gamma_b)}{3 \sigma}
\end{equation}
For $S = 0.8$, the total amplified field---accounting for the simultaneous occurrence of cavities and NRI modes at later times, each occupying half of the volume---is expected to reach
$B_\perp / B_0\sim 3.1$, which is consistent with our 3D results.
Our results then demonstrate that, with the development of the cavitation instability, a current asymmetry is generated in the beam, which allows the NRI to grow and amplify the field.

\smallskip
\noindent\textbf{\textit{Application to X-ray filaments.---}} 
We now assess whether the cavitation instability is expected to grow for the conditions of pulsar X-ray filaments. Consider a pulsar of period $P\simeq 0.1\,{\rm s}$  moving through the ISM. The pair density $n_b$ in the flux tube connected to the PWN can be estimated as follows. The so-called Goldreich–Julian rate of pair production is $\dot{N}_{\rm GJ}\simeq 2\times10^{32}(P/0.1\,{\rm s})^{-2}\,{\rm s}^{-1}$, and a realistic pulsar will produce $\dot{N}_{\rm tot}=\kappa \dot{N}_{\rm GJ}$ pairs, where $\kappa\sim 10^5$ is the multiplicity. Let us assume that the PWN particle spectrum is a broken power law with a break at 500 GeV and indices 1.4 (below) and 2.1 (above) \cite{torres+14,orusa+25}. For this spectrum, the injection rate of particles whose energy exceeds 1 TeV is $0.006\, \dot{N}_{\rm tot}$, and their mean energy is $\simeq 5\, {\rm TeV}$. Let us assume that particles above TeV energies are channeled into a cylindrical flux tube of radius $R=1$~pc, yielding a beam energy density (normalized to the rest mass energy density of ISM electrons) $\gamma_b\alpha\sim 10^{-6}$. This should be compared to the magnetization of background electrons, which for typical ISM velocities $v_A\simeq 3$ km/s is $\sigma\sim 2\times 10^{-7}$. While both $\gamma_b\alpha$ and $\sigma$ are much lower than our simulations can capture, the ratio $\gamma_b\alpha/(\sigma/2)\sim 10$ is exactly the same as our reference run (note, however, that it could be much lower for, e.g., larger cross-sectional radii of the flux tube). Given that this ratio controls both the growth of the cavitation instability (see also End Matter) and the saturation value of the NRI, our simulation results can be reliably applied to pulsar X-ray filaments---the only difference being the rate $\Gamma_{\rm max}\propto \alpha$ at which both instabilities grow. We therefore argue that the fate of the pair beam will be determined by the ratio $\alpha \gamma_b/\sigma$: if $\alpha \gamma_b/\sigma\gg1$, the cavitation instability will grow and seed the NRI, resulting in strong field amplification and efficient pair confinement, as envisioned by \cite{olmi+24}; or, if $\alpha \gamma_b/\sigma\ll1$, neither the cavitation nor the NRI will grow, resulting in negligible field amplification. The latter would be consistent with the high degree of X-ray polarization measured by \cite{dinsmore+25}. The cavitation instability alone is not expected to produce the magnetic field amplification required for X-ray filaments, as the fraction of beam energy converted into magnetic energy decreases when $\gamma_b \gtrsim m_i/m_e$ \cite{peterson+22}; yet we find that even when $\gamma_b \gtrsim m_i/m_e$ the degree of charge asymmetry remains large ($S \simeq 83\%$). In realistic systems, the primary role of the cavitation instability is to seed the conditions for the growth of the NRI.

\smallskip
\noindent\textbf{\textit{Conclusions.---}} 
We have presented 2D and 3D PIC simulations of the pair-driven cavitation instability in a magnetized electron--proton plasma, motivated by the mysterious suppression of particle diffusion inferred from X-ray filaments of bow-shock pulsar wind nebulae. Our results demonstrate that the cavitation instability, originally discussed in unmagnetized setups \cite{peterson+21,peterson+22}, can grow if the energy density of the pair beam is larger than the combined thermal and magnetic energy densities of the background. In such cases, the instability produces strong, large-scale magnetized cavities that confine the beam electrons, while beam positrons remain unconfined. The resulting net (positron) current drives the NRI, further amplifying the magnetic field. In short, the early growth of the cavitation instability is essential in mediating the onset of the Bell instability at later times, in a configuration that would not otherwise permit NRI growth. Our results provide a kinetic foundation for understanding the multi-scale environments of pulsar wind nebulae and their role in self-confinement of energetic pulsar pairs.

\begin{comment}
\begin{itemize}
    \item Cavitation, assuming filaments lenght is 10 times the transverse size $\lambda_B$=$10\sqrt{\frac{8}{\alpha}{\rm min}(\gamma_b,m_i/m_e)} \, d_e$
    \item Bell: $\lambda_{\rm fast}$=$2 \frac{\sqrt{\sigma}}{S \alpha} \, d_e$   
    \item Larmor radius= $\frac{\gamma_b}{\sqrt{\sigma}} \, d_e$
\end{itemize}
\end{comment}

\begin{acknowledgments}
\smallskip
We thank Siddhartha Gupta, Barbara Olmi, Elena Amato, Taiki Jikei, Damiano Caprioli, Anatoly Spitkosvky, Seth Gagnon, Oleg Kargaltsev and Alexander Philippov for insightful discussions. 
Simulations were performed on computational resources provided by the Princeton Research Computing and by the Flatiron Institute.
L.O. acknowledges the support of the Multimessenger Plasma Physics Center (MPPC), NSF grants PHY2206607 and PHY2206609. 
L.S. was supported by NSF grant PHY2409223. The work was supported by a grant from the Simons Foundation (MP-SCMPS-0000147, to L.S.). L.S. also acknowledges support from DoE Early Career Award DE-SC0023015.
\end{acknowledgments}

\bibliography{sample7}

\begin{thebibliography}{52}%
\makeatletter
\providecommand \@ifxundefined [1]{%
 \@ifx{#1\undefined}
}%
\providecommand \@ifnum [1]{%
 \ifnum #1\expandafter \@firstoftwo
 \else \expandafter \@secondoftwo
 \fi
}%
\providecommand \@ifx [1]{%
 \ifx #1\expandafter \@firstoftwo
 \else \expandafter \@secondoftwo
 \fi
}%
\providecommand \natexlab [1]{#1}%
\providecommand \enquote  [1]{``#1''}%
\providecommand \bibnamefont  [1]{#1}%
\providecommand \bibfnamefont [1]{#1}%
\providecommand \citenamefont [1]{#1}%
\providecommand \href@noop [0]{\@secondoftwo}%
\providecommand \href [0]{\begingroup \@sanitize@url \@href}%
\providecommand \@href[1]{\@@startlink{#1}\@@href}%
\providecommand \@@href[1]{\endgroup#1\@@endlink}%
\providecommand \@sanitize@url [0]{\catcode `\\12\catcode `\$12\catcode `\&12\catcode `\#12\catcode `\^12\catcode `\_12\catcode `\%12\relax}%
\providecommand \@@startlink[1]{}%
\providecommand \@@endlink[0]{}%
\providecommand \url  [0]{\begingroup\@sanitize@url \@url }%
\providecommand \@url [1]{\endgroup\@href {#1}{\urlprefix }}%
\providecommand \urlprefix  [0]{URL }%
\providecommand \Eprint [0]{\href }%
\providecommand \doibase [0]{http://dx.doi.org/}%
\providecommand \selectlanguage [0]{\@gobble}%
\providecommand \bibinfo  [0]{\@secondoftwo}%
\providecommand \bibfield  [0]{\@secondoftwo}%
\providecommand \translation [1]{[#1]}%
\providecommand \BibitemOpen [0]{}%
\providecommand \bibitemStop [0]{}%
\providecommand \bibitemNoStop [0]{.\EOS\space}%
\providecommand \EOS [0]{\spacefactor3000\relax}%
\providecommand \BibitemShut  [1]{\csname bibitem#1\endcsname}%
\let\auto@bib@innerbib\@empty
%</preamble>
\bibitem [{\citenamefont {{Hui}}\ and\ \citenamefont {{Becker}}(2007)}]{Hui_Becker:2007}%
  \BibitemOpen
  \bibfield  {author} {\bibinfo {author} {\bibfnamefont {C.~Y.}\ \bibnamefont {{Hui}}}\ and\ \bibinfo {author} {\bibfnamefont {W.}~\bibnamefont {{Becker}}},\ }\href {\doibase 10.1051/0004-6361:20066562} {\bibfield  {journal} {\bibinfo  {journal} {\aap}\ }\textbf {\bibinfo {volume} {467}},\ \bibinfo {pages} {1209} (\bibinfo {year} {2007})},\ \Eprint {http://arxiv.org/abs/astro-ph/0610505} {arXiv:astro-ph/0610505 [astro-ph]} \BibitemShut {NoStop}%
\bibitem [{\citenamefont {{Pavan}}\ \emph {et~al.}(2014)\citenamefont {{Pavan}}, \citenamefont {{Bordas}}, \citenamefont {{P{\"u}hlhofer}}, \citenamefont {{Filipovi{\'c}}}, \citenamefont {{De Horta}}, \citenamefont {{O'Brien}}, \citenamefont {{Balbo}}, \citenamefont {{Walter}}, \citenamefont {{Bozzo}}, \citenamefont {{Ferrigno}}, \citenamefont {{Crawford}},\ and\ \citenamefont {{Stella}}}]{Pavan:2014}%
  \BibitemOpen
  \bibfield  {author} {\bibinfo {author} {\bibfnamefont {L.}~\bibnamefont {{Pavan}}}, \bibinfo {author} {\bibfnamefont {P.}~\bibnamefont {{Bordas}}}, \bibinfo {author} {\bibfnamefont {G.}~\bibnamefont {{P{\"u}hlhofer}}}, \bibinfo {author} {\bibfnamefont {M.~D.}\ \bibnamefont {{Filipovi{\'c}}}}, \bibinfo {author} {\bibfnamefont {A.}~\bibnamefont {{De Horta}}}, \bibinfo {author} {\bibfnamefont {A.}~\bibnamefont {{O'Brien}}}, \bibinfo {author} {\bibfnamefont {M.}~\bibnamefont {{Balbo}}}, \bibinfo {author} {\bibfnamefont {R.}~\bibnamefont {{Walter}}}, \bibinfo {author} {\bibfnamefont {E.}~\bibnamefont {{Bozzo}}}, \bibinfo {author} {\bibfnamefont {C.}~\bibnamefont {{Ferrigno}}}, \bibinfo {author} {\bibfnamefont {E.}~\bibnamefont {{Crawford}}}, \ and\ \bibinfo {author} {\bibfnamefont {L.}~\bibnamefont {{Stella}}},\ }\href {\doibase 10.1051/0004-6361/201322588} {\bibfield  {journal} {\bibinfo  {journal} {\aap}\ }\textbf {\bibinfo {volume} {562}},\ \bibinfo {eid} {A122} (\bibinfo {year} {2014})},\ \Eprint
  {http://arxiv.org/abs/1309.6792} {arXiv:1309.6792 [astro-ph.HE]} \BibitemShut {NoStop}%
\bibitem [{\citenamefont {{Temim}}\ \emph {et~al.}(2015)\citenamefont {{Temim}}, \citenamefont {{Slane}}, \citenamefont {{Kolb}}, \citenamefont {{Blondin}}, \citenamefont {{Hughes}},\ and\ \citenamefont {{Bucciantini}}}]{Temim:2015}%
  \BibitemOpen
  \bibfield  {author} {\bibinfo {author} {\bibfnamefont {T.}~\bibnamefont {{Temim}}}, \bibinfo {author} {\bibfnamefont {P.}~\bibnamefont {{Slane}}}, \bibinfo {author} {\bibfnamefont {C.}~\bibnamefont {{Kolb}}}, \bibinfo {author} {\bibfnamefont {J.}~\bibnamefont {{Blondin}}}, \bibinfo {author} {\bibfnamefont {J.~P.}\ \bibnamefont {{Hughes}}}, \ and\ \bibinfo {author} {\bibfnamefont {N.}~\bibnamefont {{Bucciantini}}},\ }\href {\doibase 10.1088/0004-637X/808/1/100} {\bibfield  {journal} {\bibinfo  {journal} {\apj}\ }\textbf {\bibinfo {volume} {808}},\ \bibinfo {eid} {100} (\bibinfo {year} {2015})},\ \Eprint {http://arxiv.org/abs/1506.03069} {arXiv:1506.03069 [astro-ph.HE]} \BibitemShut {NoStop}%
\bibitem [{\citenamefont {{Klingler}}\ \emph {et~al.}(2016)\citenamefont {{Klingler}}, \citenamefont {{Kargaltsev}}, \citenamefont {{Rangelov}}, \citenamefont {{Pavlov}}, \citenamefont {{Posselt}},\ and\ \citenamefont {{Ng}}}]{Klingler:2016}%
  \BibitemOpen
  \bibfield  {author} {\bibinfo {author} {\bibfnamefont {N.}~\bibnamefont {{Klingler}}}, \bibinfo {author} {\bibfnamefont {O.}~\bibnamefont {{Kargaltsev}}}, \bibinfo {author} {\bibfnamefont {B.}~\bibnamefont {{Rangelov}}}, \bibinfo {author} {\bibfnamefont {G.~G.}\ \bibnamefont {{Pavlov}}}, \bibinfo {author} {\bibfnamefont {B.}~\bibnamefont {{Posselt}}}, \ and\ \bibinfo {author} {\bibfnamefont {C.~Y.}\ \bibnamefont {{Ng}}},\ }\href {\doibase 10.3847/0004-637X/828/2/70} {\bibfield  {journal} {\bibinfo  {journal} {\apj}\ }\textbf {\bibinfo {volume} {828}},\ \bibinfo {eid} {70} (\bibinfo {year} {2016})},\ \Eprint {http://arxiv.org/abs/1601.07174} {arXiv:1601.07174 [astro-ph.HE]} \BibitemShut {NoStop}%
\bibitem [{\citenamefont {{Klingler}}\ \emph {et~al.}(2018)\citenamefont {{Klingler}}, \citenamefont {{Kargaltsev}}, \citenamefont {{Pavlov}},\ and\ \citenamefont {{Posselt}}}]{Klingler:2018}%
  \BibitemOpen
  \bibfield  {author} {\bibinfo {author} {\bibfnamefont {N.}~\bibnamefont {{Klingler}}}, \bibinfo {author} {\bibfnamefont {O.}~\bibnamefont {{Kargaltsev}}}, \bibinfo {author} {\bibfnamefont {G.~G.}\ \bibnamefont {{Pavlov}}}, \ and\ \bibinfo {author} {\bibfnamefont {B.}~\bibnamefont {{Posselt}}},\ }\href {\doibase 10.3847/1538-4357/aae0f1} {\bibfield  {journal} {\bibinfo  {journal} {\apj}\ }\textbf {\bibinfo {volume} {868}},\ \bibinfo {eid} {119} (\bibinfo {year} {2018})},\ \Eprint {http://arxiv.org/abs/1809.04664} {arXiv:1809.04664 [astro-ph.HE]} \BibitemShut {NoStop}%
\bibitem [{\citenamefont {{Medvedev}}\ \emph {et~al.}(2019)\citenamefont {{Medvedev}}, \citenamefont {{Karpova}}, \citenamefont {{Shibanov}}, \citenamefont {{Zyuzin}},\ and\ \citenamefont {{Pavlov}}}]{Medvedev:2019}%
  \BibitemOpen
  \bibfield  {author} {\bibinfo {author} {\bibfnamefont {O.~D.}\ \bibnamefont {{Medvedev}}}, \bibinfo {author} {\bibfnamefont {A.~V.}\ \bibnamefont {{Karpova}}}, \bibinfo {author} {\bibfnamefont {Y.~A.}\ \bibnamefont {{Shibanov}}}, \bibinfo {author} {\bibfnamefont {D.~A.}\ \bibnamefont {{Zyuzin}}}, \ and\ \bibinfo {author} {\bibfnamefont {G.~G.}\ \bibnamefont {{Pavlov}}},\ }in\ \href {\doibase 10.1088/1742-6596/1400/2/022018} {\emph {\bibinfo {booktitle} {Journal of Physics Conference Series}}},\ \bibinfo {series} {Journal of Physics Conference Series}, Vol.\ \bibinfo {volume} {1400}\ (\bibinfo {year} {2019})\ p.\ \bibinfo {pages} {022018},\ \Eprint {http://arxiv.org/abs/2002.11958} {arXiv:2002.11958 [astro-ph.HE]} \BibitemShut {NoStop}%
\bibitem [{\citenamefont {{Marelli}}\ \emph {et~al.}(2019)\citenamefont {{Marelli}}, \citenamefont {{Tiengo}}, \citenamefont {{De Luca}}, \citenamefont {{Mignani}}, \citenamefont {{Salvetti}}, \citenamefont {{Saz Parkinson}},\ and\ \citenamefont {{Lisini}}}]{Marelli:2019}%
  \BibitemOpen
  \bibfield  {author} {\bibinfo {author} {\bibfnamefont {M.}~\bibnamefont {{Marelli}}}, \bibinfo {author} {\bibfnamefont {A.}~\bibnamefont {{Tiengo}}}, \bibinfo {author} {\bibfnamefont {A.}~\bibnamefont {{De Luca}}}, \bibinfo {author} {\bibfnamefont {R.~P.}\ \bibnamefont {{Mignani}}}, \bibinfo {author} {\bibfnamefont {D.}~\bibnamefont {{Salvetti}}}, \bibinfo {author} {\bibfnamefont {P.~M.}\ \bibnamefont {{Saz Parkinson}}}, \ and\ \bibinfo {author} {\bibfnamefont {G.}~\bibnamefont {{Lisini}}},\ }\href {\doibase 10.1051/0004-6361/201833464} {\bibfield  {journal} {\bibinfo  {journal} {\aap}\ }\textbf {\bibinfo {volume} {624}},\ \bibinfo {eid} {A53} (\bibinfo {year} {2019})},\ \Eprint {http://arxiv.org/abs/1808.06966} {arXiv:1808.06966 [astro-ph.HE]} \BibitemShut {NoStop}%
\bibitem [{\citenamefont {{Bordas}}\ and\ \citenamefont {{Zhang}}(2020)}]{Bordas_Zhang:2020}%
  \BibitemOpen
  \bibfield  {author} {\bibinfo {author} {\bibfnamefont {P.}~\bibnamefont {{Bordas}}}\ and\ \bibinfo {author} {\bibfnamefont {X.}~\bibnamefont {{Zhang}}},\ }\href {\doibase 10.1051/0004-6361/202039327} {\bibfield  {journal} {\bibinfo  {journal} {\aap}\ }\textbf {\bibinfo {volume} {644}},\ \bibinfo {eid} {L4} (\bibinfo {year} {2020})},\ \Eprint {http://arxiv.org/abs/2011.08593} {arXiv:2011.08593 [astro-ph.HE]} \BibitemShut {NoStop}%
\bibitem [{\citenamefont {{Zhang}}\ \emph {et~al.}(2020)\citenamefont {{Zhang}}, \citenamefont {{Zhu}}, \citenamefont {{Li}}, \citenamefont {{Pasham}}, \citenamefont {{Li}}, \citenamefont {{Clavel}}, \citenamefont {{Baganoff}}, \citenamefont {{Perez}}, \citenamefont {{Mori}},\ and\ \citenamefont {{Hailey}}}]{Zhang:2020}%
  \BibitemOpen
  \bibfield  {author} {\bibinfo {author} {\bibfnamefont {S.}~\bibnamefont {{Zhang}}}, \bibinfo {author} {\bibfnamefont {Z.}~\bibnamefont {{Zhu}}}, \bibinfo {author} {\bibfnamefont {H.}~\bibnamefont {{Li}}}, \bibinfo {author} {\bibfnamefont {D.}~\bibnamefont {{Pasham}}}, \bibinfo {author} {\bibfnamefont {Z.}~\bibnamefont {{Li}}}, \bibinfo {author} {\bibfnamefont {M.}~\bibnamefont {{Clavel}}}, \bibinfo {author} {\bibfnamefont {F.~K.}\ \bibnamefont {{Baganoff}}}, \bibinfo {author} {\bibfnamefont {K.}~\bibnamefont {{Perez}}}, \bibinfo {author} {\bibfnamefont {K.}~\bibnamefont {{Mori}}}, \ and\ \bibinfo {author} {\bibfnamefont {C.~J.}\ \bibnamefont {{Hailey}}},\ }\href {\doibase 10.3847/1538-4357/ab7dc1} {\bibfield  {journal} {\bibinfo  {journal} {\apj}\ }\textbf {\bibinfo {volume} {893}},\ \bibinfo {eid} {3} (\bibinfo {year} {2020})},\ \Eprint {http://arxiv.org/abs/2003.03453} {arXiv:2003.03453 [astro-ph.HE]} \BibitemShut {NoStop}%
\bibitem [{\citenamefont {{Klingler}}\ \emph {et~al.}(2020)\citenamefont {{Klingler}}, \citenamefont {{Yang}}, \citenamefont {{Hare}}, \citenamefont {{Kargaltsev}}, \citenamefont {{Pavlov}},\ and\ \citenamefont {{Posselt}}}]{Klingler_Yang:2020}%
  \BibitemOpen
  \bibfield  {author} {\bibinfo {author} {\bibfnamefont {N.}~\bibnamefont {{Klingler}}}, \bibinfo {author} {\bibfnamefont {H.}~\bibnamefont {{Yang}}}, \bibinfo {author} {\bibfnamefont {J.}~\bibnamefont {{Hare}}}, \bibinfo {author} {\bibfnamefont {O.}~\bibnamefont {{Kargaltsev}}}, \bibinfo {author} {\bibfnamefont {G.~G.}\ \bibnamefont {{Pavlov}}}, \ and\ \bibinfo {author} {\bibfnamefont {B.}~\bibnamefont {{Posselt}}},\ }\href {\doibase 10.3847/1538-4357/abaf4b} {\bibfield  {journal} {\bibinfo  {journal} {\apj}\ }\textbf {\bibinfo {volume} {901}},\ \bibinfo {eid} {157} (\bibinfo {year} {2020})},\ \Eprint {http://arxiv.org/abs/2008.09200} {arXiv:2008.09200 [astro-ph.HE]} \BibitemShut {NoStop}%
\bibitem [{\citenamefont {{de Vries}}\ and\ \citenamefont {{Romani}}(2022)}]{deVries-J2030:2022}%
  \BibitemOpen
  \bibfield  {author} {\bibinfo {author} {\bibfnamefont {M.}~\bibnamefont {{de Vries}}}\ and\ \bibinfo {author} {\bibfnamefont {R.~W.}\ \bibnamefont {{Romani}}},\ }\href {\doibase 10.3847/1538-4357/ac5739} {\bibfield  {journal} {\bibinfo  {journal} {\apj}\ }\textbf {\bibinfo {volume} {928}},\ \bibinfo {eid} {39} (\bibinfo {year} {2022})},\ \Eprint {http://arxiv.org/abs/2202.03506} {arXiv:2202.03506 [astro-ph.HE]} \BibitemShut {NoStop}%
\bibitem [{\citenamefont {Dinsmore}\ and\ \citenamefont {Romani}(2024)}]{dinsmore+24}%
  \BibitemOpen
  \bibfield  {author} {\bibinfo {author} {\bibfnamefont {J.~T.}\ \bibnamefont {Dinsmore}}\ and\ \bibinfo {author} {\bibfnamefont {R.~W.}\ \bibnamefont {Romani}},\ }\href {\doibase 10.3847/1538-4357/ad8344} {\bibfield  {journal} {\bibinfo  {journal} {The Astrophysical Journal}\ }\textbf {\bibinfo {volume} {976}},\ \bibinfo {pages} {4} (\bibinfo {year} {2024})}\BibitemShut {NoStop}%
\bibitem [{\citenamefont {{de Vries}}\ \emph {et~al.}(2022)\citenamefont {{de Vries}}, \citenamefont {{Romani}}, \citenamefont {{Kargaltsev}}, \citenamefont {{Pavlov}}, \citenamefont {{Posselt}}, \citenamefont {{Slane}}, \citenamefont {{Bucciantini}}, \citenamefont {{Ng}},\ and\ \citenamefont {{Klingler}}}]{deVries:2022}%
  \BibitemOpen
  \bibfield  {author} {\bibinfo {author} {\bibfnamefont {M.}~\bibnamefont {{de Vries}}}, \bibinfo {author} {\bibfnamefont {R.~W.}\ \bibnamefont {{Romani}}}, \bibinfo {author} {\bibfnamefont {O.}~\bibnamefont {{Kargaltsev}}}, \bibinfo {author} {\bibfnamefont {G.}~\bibnamefont {{Pavlov}}}, \bibinfo {author} {\bibfnamefont {B.}~\bibnamefont {{Posselt}}}, \bibinfo {author} {\bibfnamefont {P.}~\bibnamefont {{Slane}}}, \bibinfo {author} {\bibfnamefont {N.}~\bibnamefont {{Bucciantini}}}, \bibinfo {author} {\bibfnamefont {C.~Y.}\ \bibnamefont {{Ng}}}, \ and\ \bibinfo {author} {\bibfnamefont {N.}~\bibnamefont {{Klingler}}},\ }\href {\doibase 10.3847/1538-4357/ac9794} {\bibfield  {journal} {\bibinfo  {journal} {\apj}\ }\textbf {\bibinfo {volume} {939}},\ \bibinfo {eid} {70} (\bibinfo {year} {2022})},\ \Eprint {http://arxiv.org/abs/2210.01228} {arXiv:2210.01228 [astro-ph.HE]} \BibitemShut {NoStop}%
\bibitem [{\citenamefont {Dinsmore}\ \emph {et~al.}(2025)\citenamefont {Dinsmore}, \citenamefont {Romani}, \citenamefont {Mandarakas}, \citenamefont {Blinov},\ and\ \citenamefont {Liodakis}}]{dinsmore+25}%
  \BibitemOpen
  \bibfield  {author} {\bibinfo {author} {\bibfnamefont {J.~T.}\ \bibnamefont {Dinsmore}}, \bibinfo {author} {\bibfnamefont {R.~W.}\ \bibnamefont {Romani}}, \bibinfo {author} {\bibfnamefont {N.}~\bibnamefont {Mandarakas}}, \bibinfo {author} {\bibfnamefont {D.}~\bibnamefont {Blinov}}, \ and\ \bibinfo {author} {\bibfnamefont {I.}~\bibnamefont {Liodakis}},\ }\href {\doibase 10.3847/1538-4357/adafa8} {\bibfield  {journal} {\bibinfo  {journal} {The Astrophysical Journal}\ }\textbf {\bibinfo {volume} {980}},\ \bibinfo {pages} {229} (\bibinfo {year} {2025})}\BibitemShut {NoStop}%
\bibitem [{\citenamefont {{Pavan}}\ \emph {et~al.}(2016)\citenamefont {{Pavan}}, \citenamefont {{P{\"u}hlhofer}}, \citenamefont {{Bordas}}, \citenamefont {{Audard}}, \citenamefont {{Balbo}}, \citenamefont {{Bozzo}}, \citenamefont {{Eckert}}, \citenamefont {{Ferrigno}}, \citenamefont {{Filipovi{\'c}}}, \citenamefont {{Verdugo}},\ and\ \citenamefont {{Walter}}}]{Pavan:2016}%
  \BibitemOpen
  \bibfield  {author} {\bibinfo {author} {\bibfnamefont {L.}~\bibnamefont {{Pavan}}}, \bibinfo {author} {\bibfnamefont {G.}~\bibnamefont {{P{\"u}hlhofer}}}, \bibinfo {author} {\bibfnamefont {P.}~\bibnamefont {{Bordas}}}, \bibinfo {author} {\bibfnamefont {M.}~\bibnamefont {{Audard}}}, \bibinfo {author} {\bibfnamefont {M.}~\bibnamefont {{Balbo}}}, \bibinfo {author} {\bibfnamefont {E.}~\bibnamefont {{Bozzo}}}, \bibinfo {author} {\bibfnamefont {D.}~\bibnamefont {{Eckert}}}, \bibinfo {author} {\bibfnamefont {C.}~\bibnamefont {{Ferrigno}}}, \bibinfo {author} {\bibfnamefont {M.~D.}\ \bibnamefont {{Filipovi{\'c}}}}, \bibinfo {author} {\bibfnamefont {M.}~\bibnamefont {{Verdugo}}}, \ and\ \bibinfo {author} {\bibfnamefont {R.}~\bibnamefont {{Walter}}},\ }\href {\doibase 10.1051/0004-6361/201527703} {\bibfield  {journal} {\bibinfo  {journal} {\aap}\ }\textbf {\bibinfo {volume} {591}},\ \bibinfo {eid} {A91} (\bibinfo {year} {2016})},\ \Eprint {http://arxiv.org/abs/1511.01944} {arXiv:1511.01944 [astro-ph.HE]} \BibitemShut
  {NoStop}%
\bibitem [{\citenamefont {{Klingler}}\ \emph {et~al.}(2023)\citenamefont {{Klingler}}, \citenamefont {{Hare}}, \citenamefont {{Kargaltsev}}, \citenamefont {{Pavlov}},\ and\ \citenamefont {{Tomsick}}}]{Klingler:2023}%
  \BibitemOpen
  \bibfield  {author} {\bibinfo {author} {\bibfnamefont {N.}~\bibnamefont {{Klingler}}}, \bibinfo {author} {\bibfnamefont {J.}~\bibnamefont {{Hare}}}, \bibinfo {author} {\bibfnamefont {O.}~\bibnamefont {{Kargaltsev}}}, \bibinfo {author} {\bibfnamefont {G.~G.}\ \bibnamefont {{Pavlov}}}, \ and\ \bibinfo {author} {\bibfnamefont {J.}~\bibnamefont {{Tomsick}}},\ }\href {\doibase 10.3847/1538-4357/accd60} {\bibfield  {journal} {\bibinfo  {journal} {\apj}\ }\textbf {\bibinfo {volume} {950}},\ \bibinfo {eid} {177} (\bibinfo {year} {2023})},\ \Eprint {http://arxiv.org/abs/2212.03952} {arXiv:2212.03952 [astro-ph.HE]} \BibitemShut {NoStop}%
\bibitem [{\citenamefont {{Olmi}}\ \emph {et~al.}(2024)\citenamefont {{Olmi}}, \citenamefont {{Amato}}, \citenamefont {{Bandiera}},\ and\ \citenamefont {{Blasi}}}]{olmi+24}%
  \BibitemOpen
  \bibfield  {author} {\bibinfo {author} {\bibfnamefont {B.}~\bibnamefont {{Olmi}}}, \bibinfo {author} {\bibfnamefont {E.}~\bibnamefont {{Amato}}}, \bibinfo {author} {\bibfnamefont {R.}~\bibnamefont {{Bandiera}}}, \ and\ \bibinfo {author} {\bibfnamefont {P.}~\bibnamefont {{Blasi}}},\ }\href {\doibase 10.1051/0004-6361/202449382} {\bibfield  {journal} {\bibinfo  {journal} {\aap}\ }\textbf {\bibinfo {volume} {684}},\ \bibinfo {eid} {L1} (\bibinfo {year} {2024})},\ \Eprint {http://arxiv.org/abs/2403.03616} {arXiv:2403.03616 [astro-ph.HE]} \BibitemShut {NoStop}%
\bibitem [{\citenamefont {{Bell}}(2004)}]{Bell:2004}%
  \BibitemOpen
  \bibfield  {author} {\bibinfo {author} {\bibfnamefont {A.~R.}\ \bibnamefont {{Bell}}},\ }\href {\doibase 10.1111/j.1365-2966.2004.08097.x} {\bibfield  {journal} {\bibinfo  {journal} {\mnras}\ }\textbf {\bibinfo {volume} {353}},\ \bibinfo {pages} {550} (\bibinfo {year} {2004})}\BibitemShut {NoStop}%
\bibitem [{\citenamefont {Plotnikov}\ \emph {et~al.}(2024)\citenamefont {Plotnikov}, \citenamefont {van Marle}, \citenamefont {Guépin}, \citenamefont {Marcowith},\ and\ \citenamefont {Martin}}]{plotnikov+24}%
  \BibitemOpen
  \bibfield  {author} {\bibinfo {author} {\bibfnamefont {I.}~\bibnamefont {Plotnikov}}, \bibinfo {author} {\bibfnamefont {A.~J.}\ \bibnamefont {van Marle}}, \bibinfo {author} {\bibfnamefont {C.}~\bibnamefont {Guépin}}, \bibinfo {author} {\bibfnamefont {A.}~\bibnamefont {Marcowith}}, \ and\ \bibinfo {author} {\bibfnamefont {P.}~\bibnamefont {Martin}},\ }\href {\doibase 10.1051/0004-6361/202449661} {\bibfield  {journal} {\bibinfo  {journal} {Astronomy \& Astrophysics}\ }\textbf {\bibinfo {volume} {688}},\ \bibinfo {pages} {A134} (\bibinfo {year} {2024})}\BibitemShut {NoStop}%
\bibitem [{\citenamefont {{Bandiera}}(2008)}]{Bandiera:2008}%
  \BibitemOpen
  \bibfield  {author} {\bibinfo {author} {\bibfnamefont {R.}~\bibnamefont {{Bandiera}}},\ }\href {\doibase 10.1051/0004-6361:200810666} {\bibfield  {journal} {\bibinfo  {journal} {\aap}\ }\textbf {\bibinfo {volume} {490}},\ \bibinfo {pages} {L3} (\bibinfo {year} {2008})},\ \Eprint {http://arxiv.org/abs/0809.2159} {arXiv:0809.2159 [astro-ph]} \BibitemShut {NoStop}%
\bibitem [{\citenamefont {{Olmi}}\ and\ \citenamefont {{Bucciantini}}(2019)}]{Olmi_Bucciantini:2019a}%
  \BibitemOpen
  \bibfield  {author} {\bibinfo {author} {\bibfnamefont {B.}~\bibnamefont {{Olmi}}}\ and\ \bibinfo {author} {\bibfnamefont {N.}~\bibnamefont {{Bucciantini}}},\ }\href {\doibase 10.1093/mnras/stz382} {\bibfield  {journal} {\bibinfo  {journal} {\mnras}\ }\textbf {\bibinfo {volume} {484}},\ \bibinfo {pages} {5755} (\bibinfo {year} {2019})},\ \Eprint {http://arxiv.org/abs/1902.00442} {arXiv:1902.00442 [astro-ph.HE]} \BibitemShut {NoStop}%
\bibitem [{\citenamefont {{Abeysekara}}\ \emph {et~al.}(2017)\citenamefont {{Abeysekara}} \emph {et~al.}}]{hawc17}%
  \BibitemOpen
  \bibfield  {author} {\bibinfo {author} {\bibfnamefont {A.~U.}\ \bibnamefont {{Abeysekara}}} \emph {et~al.},\ }\href {\doibase 10.1126/science.aan4880} {\bibfield  {journal} {\bibinfo  {journal} {Science}\ }\textbf {\bibinfo {volume} {358}},\ \bibinfo {pages} {911} (\bibinfo {year} {2017})},\ \Eprint {http://arxiv.org/abs/1711.06223} {arXiv:1711.06223 [astro-ph.HE]} \BibitemShut {NoStop}%
\bibitem [{\citenamefont {{Albert}}\ \emph {et~al.}(2023)\citenamefont {{Albert}} \emph {et~al.}}]{hawc23}%
  \BibitemOpen
  \bibfield  {author} {\bibinfo {author} {\bibfnamefont {A.}~\bibnamefont {{Albert}}} \emph {et~al.},\ }\href {\doibase 10.3847/2041-8213/acb5ee} {\bibfield  {journal} {\bibinfo  {journal} {\apjl}\ }\textbf {\bibinfo {volume} {944}},\ \bibinfo {eid} {L29} (\bibinfo {year} {2023})},\ \Eprint {http://arxiv.org/abs/2301.04646} {arXiv:2301.04646 [astro-ph.HE]} \BibitemShut {NoStop}%
\bibitem [{\citenamefont {{Abdo}}\ \emph {et~al.}(2009)\citenamefont {{Abdo}}, \citenamefont {{Allen}}, \citenamefont {{Aune}} \emph {et~al.}}]{milagro09}%
  \BibitemOpen
  \bibfield  {author} {\bibinfo {author} {\bibfnamefont {A.~A.}\ \bibnamefont {{Abdo}}}, \bibinfo {author} {\bibfnamefont {B.~T.}\ \bibnamefont {{Allen}}}, \bibinfo {author} {\bibfnamefont {T.}~\bibnamefont {{Aune}}},  \emph {et~al.},\ }\href {\doibase 10.1088/0004-637X/700/2/L127} {\bibfield  {journal} {\bibinfo  {journal} {\apjl}\ }\textbf {\bibinfo {volume} {700}},\ \bibinfo {pages} {L127} (\bibinfo {year} {2009})},\ \Eprint {http://arxiv.org/abs/0904.1018} {arXiv:0904.1018 [astro-ph.HE]} \BibitemShut {NoStop}%
\bibitem [{\citenamefont {Aharonian}\ \emph {et~al.}(2023)\citenamefont {Aharonian} \emph {et~al.}}]{hess23}%
  \BibitemOpen
  \bibfield  {author} {\bibinfo {author} {\bibfnamefont {F.}~\bibnamefont {Aharonian}} \emph {et~al.} (\bibinfo {collaboration} {H.E.S.S.}),\ }\href {\doibase 10.1051/0004-6361/202245776} {\bibfield  {journal} {\bibinfo  {journal} {Astron. Astrophys.}\ }\textbf {\bibinfo {volume} {673}},\ \bibinfo {pages} {A148} (\bibinfo {year} {2023})},\ \Eprint {http://arxiv.org/abs/2304.02631} {arXiv:2304.02631 [astro-ph.HE]} \BibitemShut {NoStop}%
\bibitem [{\citenamefont {{Aharonian}}\ \emph {et~al.}(2021)\citenamefont {{Aharonian}} \emph {et~al.}}]{lhaaso21}%
  \BibitemOpen
  \bibfield  {author} {\bibinfo {author} {\bibfnamefont {F.}~\bibnamefont {{Aharonian}}} \emph {et~al.},\ }\href {\doibase 10.1103/PhysRevLett.126.241103} {\bibfield  {journal} {\bibinfo  {journal} {\prl}\ }\textbf {\bibinfo {volume} {126}},\ \bibinfo {eid} {241103} (\bibinfo {year} {2021})},\ \Eprint {http://arxiv.org/abs/2106.09396} {arXiv:2106.09396 [astro-ph.HE]} \BibitemShut {NoStop}%
\bibitem [{\citenamefont {Cao}\ \emph {et~al.}(2024)\citenamefont {Cao} \emph {et~al.}}]{cao+24}%
  \BibitemOpen
  \bibfield  {author} {\bibinfo {author} {\bibfnamefont {Z.}~\bibnamefont {Cao}} \emph {et~al.},\ }\href {https://arxiv.org/abs/2410.04425} {\  (\bibinfo {year} {2024})},\ \Eprint {http://arxiv.org/abs/2410.04425} {arXiv:2410.04425 [astro-ph.HE]} \BibitemShut {NoStop}%
\bibitem [{\citenamefont {Di~Mauro}\ \emph {et~al.}(2019)\citenamefont {Di~Mauro}, \citenamefont {Manconi},\ and\ \citenamefont {Donato}}]{dimauro+19}%
  \BibitemOpen
  \bibfield  {author} {\bibinfo {author} {\bibfnamefont {M.}~\bibnamefont {Di~Mauro}}, \bibinfo {author} {\bibfnamefont {S.}~\bibnamefont {Manconi}}, \ and\ \bibinfo {author} {\bibfnamefont {F.}~\bibnamefont {Donato}},\ }\href {\doibase 10.1103/PhysRevD.100.123015} {\bibfield  {journal} {\bibinfo  {journal} {Phys. Rev. D}\ }\textbf {\bibinfo {volume} {100}},\ \bibinfo {pages} {123015} (\bibinfo {year} {2019})}\BibitemShut {NoStop}%
\bibitem [{\citenamefont {Dinsmore}\ and\ \citenamefont {Romani}(2026)}]{dinsmore+26}%
  \BibitemOpen
  \bibfield  {author} {\bibinfo {author} {\bibfnamefont {J.~T.}\ \bibnamefont {Dinsmore}}\ and\ \bibinfo {author} {\bibfnamefont {R.~W.}\ \bibnamefont {Romani}},\ }\href {https://arxiv.org/abs/2603.20532} {\enquote {\bibinfo {title} {A physical model of pulsar x-ray filaments},}\ } (\bibinfo {year} {2026}),\ \Eprint {http://arxiv.org/abs/2603.20532} {arXiv:2603.20532 [astro-ph.HE]} \BibitemShut {NoStop}%
\bibitem [{\citenamefont {{Evoli}}\ \emph {et~al.}(2018)\citenamefont {{Evoli}}, \citenamefont {{Linden}},\ and\ \citenamefont {{Morlino}}}]{evoli+18}%
  \BibitemOpen
  \bibfield  {author} {\bibinfo {author} {\bibfnamefont {C.}~\bibnamefont {{Evoli}}}, \bibinfo {author} {\bibfnamefont {T.}~\bibnamefont {{Linden}}}, \ and\ \bibinfo {author} {\bibfnamefont {G.}~\bibnamefont {{Morlino}}},\ }\href {\doibase 10.1103/PhysRevD.98.063017} {\bibfield  {journal} {\bibinfo  {journal} {\prd}\ }\textbf {\bibinfo {volume} {98}},\ \bibinfo {eid} {063017} (\bibinfo {year} {2018})},\ \Eprint {http://arxiv.org/abs/1807.09263} {arXiv:1807.09263 [astro-ph.HE]} \BibitemShut {NoStop}%
\bibitem [{\citenamefont {Mukhopadhyay}\ and\ \citenamefont {Linden}(2022)}]{Mukhopadhyay22}%
  \BibitemOpen
  \bibfield  {author} {\bibinfo {author} {\bibfnamefont {P.}~\bibnamefont {Mukhopadhyay}}\ and\ \bibinfo {author} {\bibfnamefont {T.}~\bibnamefont {Linden}},\ }\href {\doibase 10.1103/PhysRevD.105.123008} {\bibfield  {journal} {\bibinfo  {journal} {Phys. Rev. D}\ }\textbf {\bibinfo {volume} {105}},\ \bibinfo {pages} {123008} (\bibinfo {year} {2022})},\ \Eprint {http://arxiv.org/abs/2111.01143} {arXiv:2111.01143 [astro-ph.HE]} \BibitemShut {NoStop}%
\bibitem [{\citenamefont {Orusa}\ \emph {et~al.}(2021)\citenamefont {Orusa}, \citenamefont {Manconi}, \citenamefont {Donato},\ and\ \citenamefont {Di~Mauro}}]{orusa+21}%
  \BibitemOpen
  \bibfield  {author} {\bibinfo {author} {\bibfnamefont {L.}~\bibnamefont {Orusa}}, \bibinfo {author} {\bibfnamefont {S.}~\bibnamefont {Manconi}}, \bibinfo {author} {\bibfnamefont {F.}~\bibnamefont {Donato}}, \ and\ \bibinfo {author} {\bibfnamefont {M.}~\bibnamefont {Di~Mauro}},\ }\href {\doibase 10.1088/1475-7516/2021/12/014} {\bibfield  {journal} {\bibinfo  {journal} {Journal of Cosmology and Astroparticle Physics}\ }\textbf {\bibinfo {volume} {2021}},\ \bibinfo {pages} {014} (\bibinfo {year} {2021})}\BibitemShut {NoStop}%
\bibitem [{\citenamefont {Orusa}\ \emph {et~al.}(2025)\citenamefont {Orusa}, \citenamefont {Manconi}, \citenamefont {Donato},\ and\ \citenamefont {Mauro}}]{orusa+25}%
  \BibitemOpen
  \bibfield  {author} {\bibinfo {author} {\bibfnamefont {L.}~\bibnamefont {Orusa}}, \bibinfo {author} {\bibfnamefont {S.}~\bibnamefont {Manconi}}, \bibinfo {author} {\bibfnamefont {F.}~\bibnamefont {Donato}}, \ and\ \bibinfo {author} {\bibfnamefont {M.~D.}\ \bibnamefont {Mauro}},\ }\href {\doibase 10.1088/1475-7516/2025/02/029} {\bibfield  {journal} {\bibinfo  {journal} {Journal of Cosmology and Astroparticle Physics}\ }\textbf {\bibinfo {volume} {2025}},\ \bibinfo {pages} {029} (\bibinfo {year} {2025})}\BibitemShut {NoStop}%
\bibitem [{\citenamefont {Liu}\ \emph {et~al.}(2019)\citenamefont {Liu}, \citenamefont {Yan},\ and\ \citenamefont {Zhang}}]{liu+19}%
  \BibitemOpen
  \bibfield  {author} {\bibinfo {author} {\bibfnamefont {R.-Y.}\ \bibnamefont {Liu}}, \bibinfo {author} {\bibfnamefont {H.}~\bibnamefont {Yan}}, \ and\ \bibinfo {author} {\bibfnamefont {H.}~\bibnamefont {Zhang}},\ }\href {\doibase 10.1103/physrevlett.123.221103} {\bibfield  {journal} {\bibinfo  {journal} {Physical Review Letters}\ }\textbf {\bibinfo {volume} {123}} (\bibinfo {year} {2019}),\ 10.1103/physrevlett.123.221103}\BibitemShut {NoStop}%
\bibitem [{\citenamefont {De~La Torre~Luque}\ \emph {et~al.}(2022)\citenamefont {De~La Torre~Luque}, \citenamefont {Fornieri},\ and\ \citenamefont {Linden}}]{DeLaTorreLuque+22}%
  \BibitemOpen
  \bibfield  {author} {\bibinfo {author} {\bibfnamefont {P.}~\bibnamefont {De~La Torre~Luque}}, \bibinfo {author} {\bibfnamefont {O.}~\bibnamefont {Fornieri}}, \ and\ \bibinfo {author} {\bibfnamefont {T.}~\bibnamefont {Linden}},\ }\href {\doibase 10.1103/PhysRevD.106.123033} {\bibfield  {journal} {\bibinfo  {journal} {Phys. Rev. D}\ }\textbf {\bibinfo {volume} {106}},\ \bibinfo {pages} {123033} (\bibinfo {year} {2022})},\ \Eprint {http://arxiv.org/abs/2205.08544} {arXiv:2205.08544 [astro-ph.HE]} \BibitemShut {NoStop}%
\bibitem [{\citenamefont {Recchia}\ \emph {et~al.}(2021)\citenamefont {Recchia}, \citenamefont {Di~Mauro}, \citenamefont {Aharonian}, \citenamefont {Orusa}, \citenamefont {Donato}, \citenamefont {Gabici},\ and\ \citenamefont {Manconi}}]{Recchia21}%
  \BibitemOpen
  \bibfield  {author} {\bibinfo {author} {\bibfnamefont {S.}~\bibnamefont {Recchia}}, \bibinfo {author} {\bibfnamefont {M.}~\bibnamefont {Di~Mauro}}, \bibinfo {author} {\bibfnamefont {F.~A.}\ \bibnamefont {Aharonian}}, \bibinfo {author} {\bibfnamefont {L.}~\bibnamefont {Orusa}}, \bibinfo {author} {\bibfnamefont {F.}~\bibnamefont {Donato}}, \bibinfo {author} {\bibfnamefont {S.}~\bibnamefont {Gabici}}, \ and\ \bibinfo {author} {\bibfnamefont {S.}~\bibnamefont {Manconi}},\ }\href {\doibase 10.1103/PhysRevD.104.123017} {\bibfield  {journal} {\bibinfo  {journal} {Phys. Rev. D}\ }\textbf {\bibinfo {volume} {104}},\ \bibinfo {pages} {123017} (\bibinfo {year} {2021})}\BibitemShut {NoStop}%
\bibitem [{\citenamefont {L\'opez-Coto}\ and\ \citenamefont {Giacinti}(2018)}]{Lopez-Coto+18}%
  \BibitemOpen
  \bibfield  {author} {\bibinfo {author} {\bibfnamefont {R.}~\bibnamefont {L\'opez-Coto}}\ and\ \bibinfo {author} {\bibfnamefont {G.}~\bibnamefont {Giacinti}},\ }\href {\doibase 10.1093/mnras/sty1821} {\bibfield  {journal} {\bibinfo  {journal} {Mon. Not. Roy. Astron. Soc.}\ }\textbf {\bibinfo {volume} {479}},\ \bibinfo {pages} {4526} (\bibinfo {year} {2018})},\ \Eprint {http://arxiv.org/abs/1712.04373} {arXiv:1712.04373 [astro-ph.HE]} \BibitemShut {NoStop}%
\bibitem [{\citenamefont {{Fang}}\ \emph {et~al.}(2019)\citenamefont {{Fang}}, \citenamefont {{Bi}},\ and\ \citenamefont {{Yin}}}]{fang+19}%
  \BibitemOpen
  \bibfield  {author} {\bibinfo {author} {\bibfnamefont {K.}~\bibnamefont {{Fang}}}, \bibinfo {author} {\bibfnamefont {X.-J.}\ \bibnamefont {{Bi}}}, \ and\ \bibinfo {author} {\bibfnamefont {P.-F.}\ \bibnamefont {{Yin}}},\ }\href {\doibase 10.1093/mnras/stz1974} {\bibfield  {journal} {\bibinfo  {journal} {\mnras}\ }\textbf {\bibinfo {volume} {488}},\ \bibinfo {pages} {4074} (\bibinfo {year} {2019})},\ \Eprint {http://arxiv.org/abs/1903.06421} {arXiv:1903.06421 [astro-ph.HE]} \BibitemShut {NoStop}%
\bibitem [{\citenamefont {Bourguinat}\ \emph {et~al.}(2025)\citenamefont {Bourguinat}, \citenamefont {Evoli}, \citenamefont {Martin},\ and\ \citenamefont {Recchia}}]{bourguinat+25}%
  \BibitemOpen
  \bibfield  {author} {\bibinfo {author} {\bibfnamefont {L.-M.}\ \bibnamefont {Bourguinat}}, \bibinfo {author} {\bibfnamefont {C.}~\bibnamefont {Evoli}}, \bibinfo {author} {\bibfnamefont {P.}~\bibnamefont {Martin}}, \ and\ \bibinfo {author} {\bibfnamefont {S.}~\bibnamefont {Recchia}},\ }\href {https://arxiv.org/abs/2507.01495} {\enquote {\bibinfo {title} {The environment of tev halo progenitors},}\ } (\bibinfo {year} {2025}),\ \Eprint {http://arxiv.org/abs/2507.01495} {arXiv:2507.01495 [astro-ph.HE]} \BibitemShut {NoStop}%
\bibitem [{\citenamefont {Peterson}\ \emph {et~al.}(2021)\citenamefont {Peterson}, \citenamefont {Glenzer},\ and\ \citenamefont {Fiuza}}]{peterson+21}%
  \BibitemOpen
  \bibfield  {author} {\bibinfo {author} {\bibfnamefont {J.}~\bibnamefont {Peterson}}, \bibinfo {author} {\bibfnamefont {S.}~\bibnamefont {Glenzer}}, \ and\ \bibinfo {author} {\bibfnamefont {F.}~\bibnamefont {Fiuza}},\ }\href {\doibase 10.1103/physrevlett.126.215101} {\bibfield  {journal} {\bibinfo  {journal} {Physical Review Letters}\ }\textbf {\bibinfo {volume} {126}} (\bibinfo {year} {2021}),\ 10.1103/physrevlett.126.215101}\BibitemShut {NoStop}%
\bibitem [{\citenamefont {Peterson}\ \emph {et~al.}(2022)\citenamefont {Peterson}, \citenamefont {Glenzer},\ and\ \citenamefont {Fiuza}}]{peterson+22}%
  \BibitemOpen
  \bibfield  {author} {\bibinfo {author} {\bibfnamefont {J.~R.}\ \bibnamefont {Peterson}}, \bibinfo {author} {\bibfnamefont {S.}~\bibnamefont {Glenzer}}, \ and\ \bibinfo {author} {\bibfnamefont {F.}~\bibnamefont {Fiuza}},\ }\href {\doibase 10.3847/2041-8213/ac44a2} {\bibfield  {journal} {\bibinfo  {journal} {The Astrophysical Journal Letters}\ }\textbf {\bibinfo {volume} {924}},\ \bibinfo {pages} {L12} (\bibinfo {year} {2022})}\BibitemShut {NoStop}%
\bibitem [{\citenamefont {Golant}\ \emph {et~al.}(2024)\citenamefont {Golant}, \citenamefont {Vanthieghem}, \citenamefont {Groselj},\ and\ \citenamefont {Sironi}}]{golant+24}%
  \BibitemOpen
  \bibfield  {author} {\bibinfo {author} {\bibfnamefont {R.}~\bibnamefont {Golant}}, \bibinfo {author} {\bibfnamefont {A.}~\bibnamefont {Vanthieghem}}, \bibinfo {author} {\bibfnamefont {D.}~\bibnamefont {Groselj}}, \ and\ \bibinfo {author} {\bibfnamefont {L.}~\bibnamefont {Sironi}},\ }\href {https://arxiv.org/abs/2410.05388} {\enquote {\bibinfo {title} {Generation of large-scale magnetic fields upstream of gamma-ray burst afterglow shocks},}\ } (\bibinfo {year} {2024}),\ \Eprint {http://arxiv.org/abs/2410.05388} {arXiv:2410.05388 [astro-ph.HE]} \BibitemShut {NoStop}%
\bibitem [{\citenamefont {Grošelj}\ \emph {et~al.}(2024)\citenamefont {Grošelj}, \citenamefont {Sironi},\ and\ \citenamefont {Spitkovsky}}]{groselj+24}%
  \BibitemOpen
  \bibfield  {author} {\bibinfo {author} {\bibfnamefont {D.}~\bibnamefont {Grošelj}}, \bibinfo {author} {\bibfnamefont {L.}~\bibnamefont {Sironi}}, \ and\ \bibinfo {author} {\bibfnamefont {A.}~\bibnamefont {Spitkovsky}},\ }\href {\doibase 10.3847/2041-8213/ad2c8c} {\bibfield  {journal} {\bibinfo  {journal} {The Astrophysical Journal Letters}\ }\textbf {\bibinfo {volume} {963}},\ \bibinfo {pages} {L44} (\bibinfo {year} {2024})}\BibitemShut {NoStop}%
\bibitem [{\citenamefont {Bresci}\ \emph {et~al.}(2022)\citenamefont {Bresci}, \citenamefont {Gremillet},\ and\ \citenamefont {Lemoine}}]{bresci+22}%
  \BibitemOpen
  \bibfield  {author} {\bibinfo {author} {\bibfnamefont {V.}~\bibnamefont {Bresci}}, \bibinfo {author} {\bibfnamefont {L.}~\bibnamefont {Gremillet}}, \ and\ \bibinfo {author} {\bibfnamefont {M.}~\bibnamefont {Lemoine}},\ }\href {\doibase 10.1103/physreve.105.035202} {\bibfield  {journal} {\bibinfo  {journal} {Physical Review E}\ }\textbf {\bibinfo {volume} {105}} (\bibinfo {year} {2022}),\ 10.1103/physreve.105.035202}\BibitemShut {NoStop}%
\bibitem [{\citenamefont {Demidov}\ \emph {et~al.}(2025)\citenamefont {Demidov}, \citenamefont {Lyubarsky},\ and\ \citenamefont {Keshet}}]{demidov+25}%
  \BibitemOpen
  \bibfield  {author} {\bibinfo {author} {\bibfnamefont {I.}~\bibnamefont {Demidov}}, \bibinfo {author} {\bibfnamefont {Y.}~\bibnamefont {Lyubarsky}}, \ and\ \bibinfo {author} {\bibfnamefont {U.}~\bibnamefont {Keshet}},\ }\href {https://arxiv.org/abs/2511.00742} {\enquote {\bibinfo {title} {Cavitation instability in unmagnetized relativistic pair shocks},}\ } (\bibinfo {year} {2025}),\ \Eprint {http://arxiv.org/abs/2511.00742} {arXiv:2511.00742 [astro-ph.HE]} \BibitemShut {NoStop}%
\bibitem [{\citenamefont {{Weibel}}(1959)}]{weibel59}%
  \BibitemOpen
  \bibfield  {author} {\bibinfo {author} {\bibfnamefont {E.~S.}\ \bibnamefont {{Weibel}}},\ }\href {\doibase 10.1103/PhysRevLett.2.83} {\bibfield  {journal} {\bibinfo  {journal} {\prl}\ }\textbf {\bibinfo {volume} {2}},\ \bibinfo {pages} {83} (\bibinfo {year} {1959})}\BibitemShut {NoStop}%
\bibitem [{\citenamefont {{Spitkovsky}}(2005)}]{spitkovsky05}%
  \BibitemOpen
  \bibfield  {author} {\bibinfo {author} {\bibfnamefont {A.}~\bibnamefont {{Spitkovsky}}},\ }in\ \href {\doibase 10.1063/1.2141897} {\emph {\bibinfo {booktitle} {Astrophysical Sources of High Energy Particles and Radiation}}},\ \bibinfo {series} {American Institute of Physics Conference Series}, Vol.\ \bibinfo {volume} {801},\ \bibinfo {editor} {edited by\ \bibinfo {editor} {\bibfnamefont {T.}~\bibnamefont {{Bulik}}}, \bibinfo {editor} {\bibfnamefont {B.}~\bibnamefont {{Rudak}}}, \ and\ \bibinfo {editor} {\bibfnamefont {G.}~\bibnamefont {{Madejski}}}}\ (\bibinfo {year} {2005})\ pp.\ \bibinfo {pages} {345--350},\ \Eprint {http://arxiv.org/abs/astro-ph/0603211} {astro-ph/0603211} \BibitemShut {NoStop}%
\bibitem [{\citenamefont {{Davidson}}\ \emph {et~al.}(1972)\citenamefont {{Davidson}}, \citenamefont {{Hammer}}, \citenamefont {{Haber}},\ and\ \citenamefont {{Wagner}}}]{davidson+72}%
  \BibitemOpen
  \bibfield  {author} {\bibinfo {author} {\bibfnamefont {R.~C.}\ \bibnamefont {{Davidson}}}, \bibinfo {author} {\bibfnamefont {D.~A.}\ \bibnamefont {{Hammer}}}, \bibinfo {author} {\bibfnamefont {I.}~\bibnamefont {{Haber}}}, \ and\ \bibinfo {author} {\bibfnamefont {C.~E.}\ \bibnamefont {{Wagner}}},\ }\href {\doibase 10.1063/1.1693910} {\bibfield  {journal} {\bibinfo  {journal} {Physics of Fluids}\ }\textbf {\bibinfo {volume} {15}},\ \bibinfo {pages} {317} (\bibinfo {year} {1972})}\BibitemShut {NoStop}%
\bibitem [{\citenamefont {{Gupta}}\ \emph {et~al.}(2021)\citenamefont {{Gupta}}, \citenamefont {{Caprioli}},\ and\ \citenamefont {{Haggerty}}}]{gupta+21}%
  \BibitemOpen
  \bibfield  {author} {\bibinfo {author} {\bibfnamefont {S.}~\bibnamefont {{Gupta}}}, \bibinfo {author} {\bibfnamefont {D.}~\bibnamefont {{Caprioli}}}, \ and\ \bibinfo {author} {\bibfnamefont {C.~C.}\ \bibnamefont {{Haggerty}}},\ }\href {\doibase 10.3847/1538-4357/ac23cf} {\bibfield  {journal} {\bibinfo  {journal} {\apj}\ }\textbf {\bibinfo {volume} {923}},\ \bibinfo {eid} {208} (\bibinfo {year} {2021})},\ \Eprint {http://arxiv.org/abs/2106.07672} {arXiv:2106.07672 [astro-ph.HE]} \BibitemShut {NoStop}%
\bibitem [{\citenamefont {Zacharegkas}\ \emph {et~al.}(2024)\citenamefont {Zacharegkas}, \citenamefont {Caprioli}, \citenamefont {Haggerty}, \citenamefont {Gupta},\ and\ \citenamefont {Schroer}}]{zacharegkas+24}%
  \BibitemOpen
  \bibfield  {author} {\bibinfo {author} {\bibfnamefont {G.}~\bibnamefont {Zacharegkas}}, \bibinfo {author} {\bibfnamefont {D.}~\bibnamefont {Caprioli}}, \bibinfo {author} {\bibfnamefont {C.}~\bibnamefont {Haggerty}}, \bibinfo {author} {\bibfnamefont {S.}~\bibnamefont {Gupta}}, \ and\ \bibinfo {author} {\bibfnamefont {B.}~\bibnamefont {Schroer}},\ }\href {https://arxiv.org/abs/2210.08072} {\enquote {\bibinfo {title} {Modeling the saturation of the bell instability using hybrid simulations},}\ } (\bibinfo {year} {2024}),\ \Eprint {http://arxiv.org/abs/2210.08072} {arXiv:2210.08072 [astro-ph.HE]} \BibitemShut {NoStop}%
\bibitem [{\citenamefont {Lichko}\ \emph {et~al.}(2025)\citenamefont {Lichko}, \citenamefont {Caprioli}, \citenamefont {Schroer},\ and\ \citenamefont {Gupta}}]{lichko+25}%
  \BibitemOpen
  \bibfield  {author} {\bibinfo {author} {\bibfnamefont {E.}~\bibnamefont {Lichko}}, \bibinfo {author} {\bibfnamefont {D.}~\bibnamefont {Caprioli}}, \bibinfo {author} {\bibfnamefont {B.}~\bibnamefont {Schroer}}, \ and\ \bibinfo {author} {\bibfnamefont {S.}~\bibnamefont {Gupta}},\ }\href {https://arxiv.org/abs/2411.05704} {\enquote {\bibinfo {title} {Understanding streaming instabilities in the limit of high cosmic ray current density},}\ } (\bibinfo {year} {2025}),\ \Eprint {http://arxiv.org/abs/2411.05704} {arXiv:2411.05704 [astro-ph.HE]} \BibitemShut {NoStop}%
\bibitem [{\citenamefont {Torres}\ \emph {et~al.}(2014)\citenamefont {Torres}, \citenamefont {Cillis}, \citenamefont {Martín},\ and\ \citenamefont {de~Oña~Wilhelmi}}]{torres+14}%
  \BibitemOpen
  \bibfield  {author} {\bibinfo {author} {\bibfnamefont {D.~F.}\ \bibnamefont {Torres}}, \bibinfo {author} {\bibfnamefont {A.}~\bibnamefont {Cillis}}, \bibinfo {author} {\bibfnamefont {J.}~\bibnamefont {Martín}}, \ and\ \bibinfo {author} {\bibfnamefont {E.}~\bibnamefont {de~Oña~Wilhelmi}},\ }\href {https://arxiv.org/abs/1402.5485} {\enquote {\bibinfo {title} {Time-dependent modeling of tev-detected, young pulsar wind nebulae},}\ } (\bibinfo {year} {2014}),\ \Eprint {http://arxiv.org/abs/1402.5485} {arXiv:1402.5485 [astro-ph.HE]} \BibitemShut {NoStop}%
\end{thebibliography}%

\appendix

\section{End Matter}\label{appendix}

\begin{figure}[h]
\begin{center}
    \includegraphics[width=0.49\textwidth,clip=true,trim= 0 1110 0 0]{"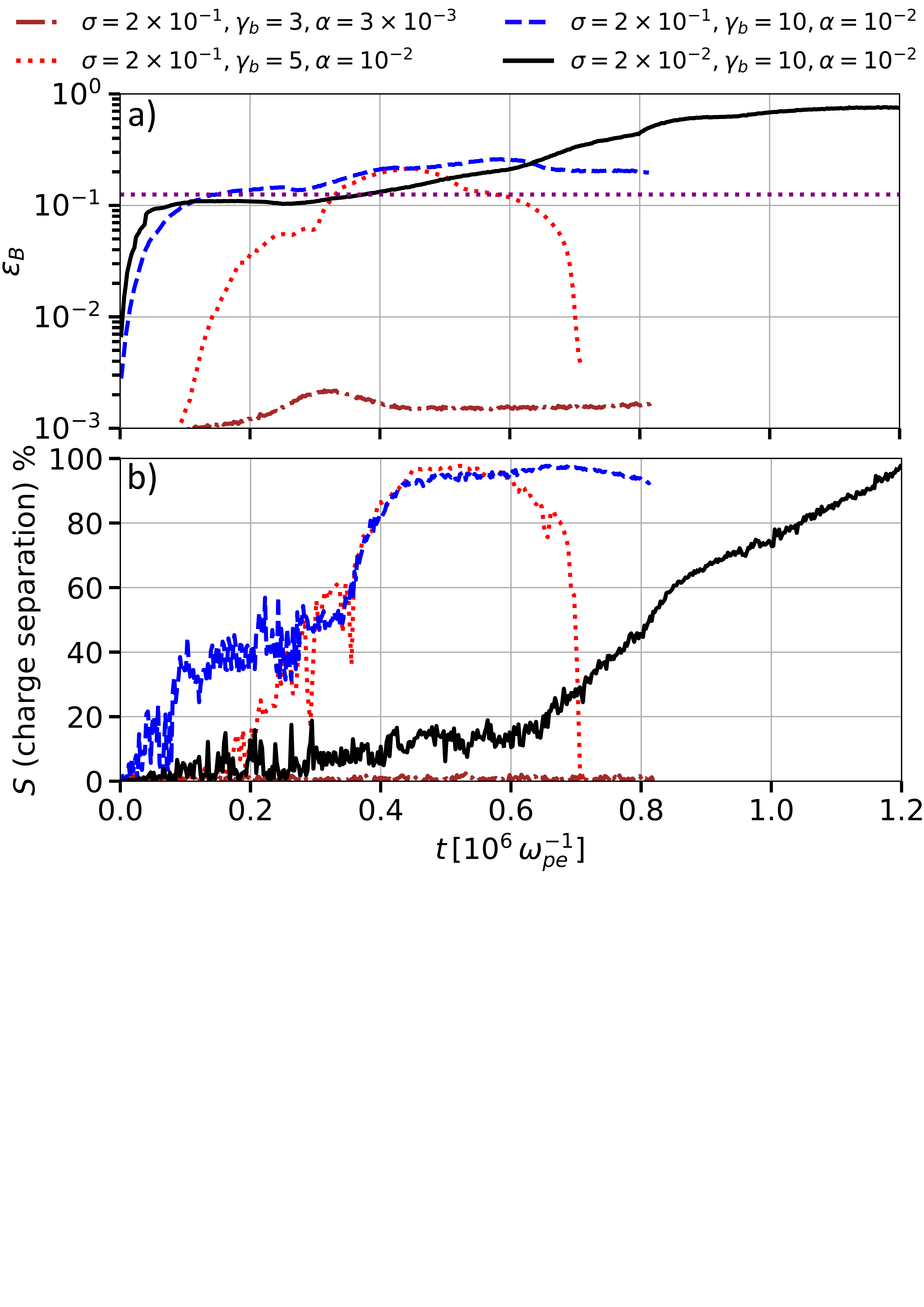"}
    \caption{Same as Figure \ref{Fig:no_refresh} but for the case in which we ``refresh'' the beam particles. We use $\Delta \gamma_b=10^{-4}$.
    } 
    \label{Fig:refresh_9000}
    \end{center}
\end{figure}

\smallskip
\noindent\textbf{\textit{Refreshed cases.---}} 
We test a different setup with respect to the case reported in the Letter, in which beam particles are stochastically reset to their initial momentum distribution over a timescale $t_{\rm refresh}$: at each timestep, beam particles are ``refreshed'' with a probability such that each particle, on average, is reset over a timescale $t_{\rm refresh}$. This mimics the injection of new PWN pairs in the plasma volume of our simulation. Note, however, that we reset the momentum, but we do not change the position of beam particles. For the refreshing process to be effective, the refresh timescale must be comparable to the decay time of the initial Weibel fields, which for the conditions explored in this work is $\sim 10^4 \, \omega_{pe}^{-1}$. We therefore test two values: an intermediate $t_{\rm refresh} = 9000 \, \omega_{pe}^{-1}$, close to the Weibel decay time, and a long $t_{\rm refresh} = 90000 \, \omega_{pe}^{-1}$, comparable to the saturation time of the cavitation instability. For $t_{\rm refresh} = 9000 \, \omega_{pe}^{-1}$, the cavitation instability develops even when the beam-to-background energy ratio is as low as $(\gamma_b \alpha/\sigma)_{\rm refresh}> 0.5$, as shown in Figure \ref{Fig:refresh_9000}a, where we plot $\epsilon_B$, i.e., the fraction of initial beam  energy  converted into magnetic energy. In contrast, the choice $t_{\rm refresh} = 90000 \, \omega_{pe}^{-1}$ does not enable the growth of the cavitation instability in cases that were  suppressed in our standard setup, but it does delay the decay of the cavitation modes when they grow (this case is not shown). 

\begin{figure}[t]
\centering    
\vspace{0.2cm}
\includegraphics[width=0.49\textwidth,clip=true,trim= 0 700 0 0]{"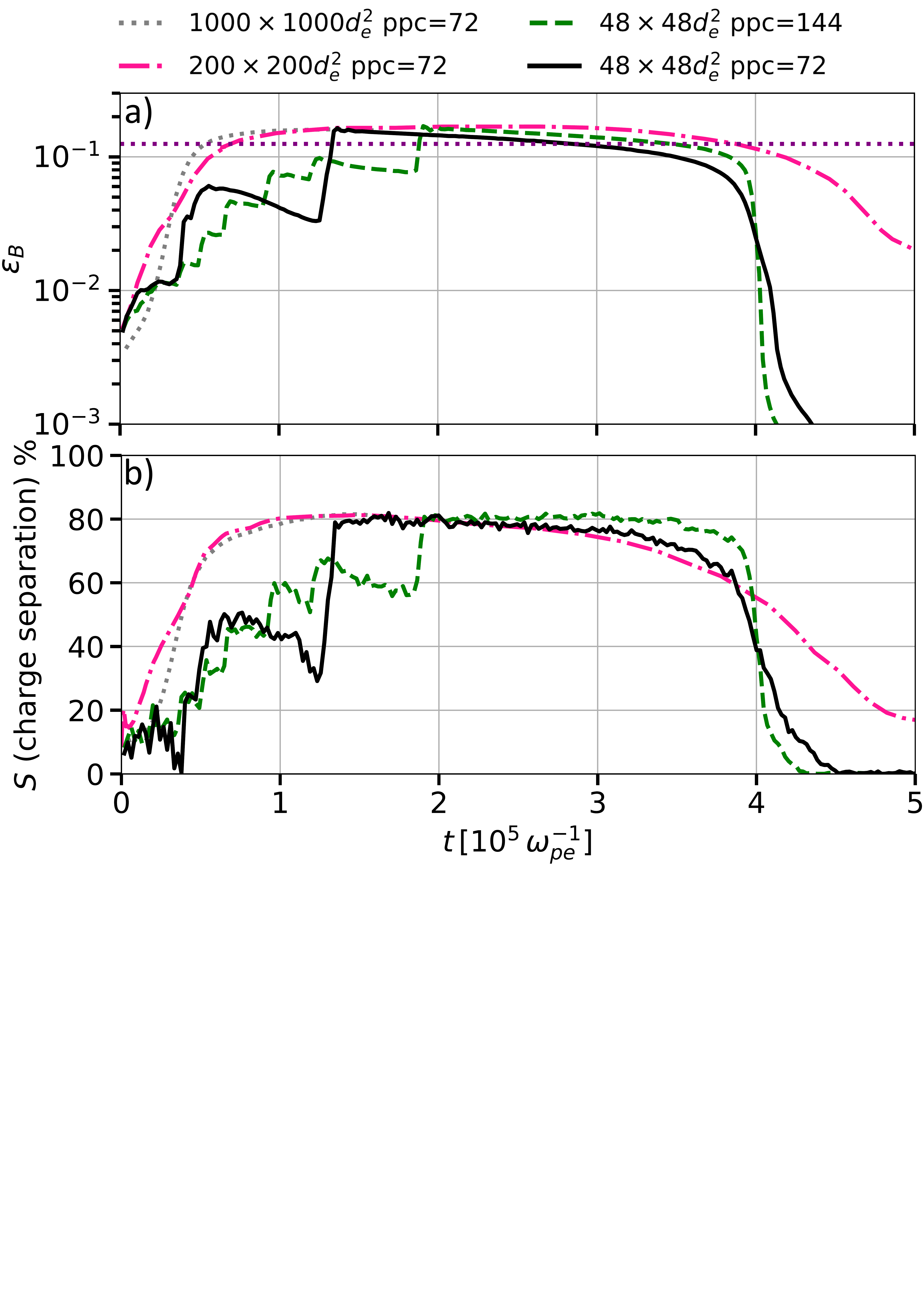"}
    \vspace{-58pt}
    \caption{Same as Figure \ref{Fig:no_refresh} of our reference case, with $\alpha \gamma_b = 10^{-1}$, $\sigma = 2\times 10^{-2}$, and $\Delta \gamma_b=10^{-4}$. The four curves correspond to different box sizes, $48 \times 48 \, d_e^2$ (solid black) and $200 \times 200 \, d_e^2$ (dotted-dashed pink) and $1000 \times 1000 \, d_e^2$ (dotted gray, until $t=2 \times 10^5 \omega_{pe}^{-1}$), and to different ppc (dashed green).
    } 
    \label{Fig:numeric}
\end{figure}

The level of magnetic field amplification for $t_{\rm refresh} = 9000 \, \omega_{pe}^{-1}$ exceeds both the non-refreshed case and theoretical predictions, due to the additional energy injected into the beam by the refreshing process. We find that $e^-$-driven cavities still live longer than $e^+$-driven cavities, although both types survive longer than in corresponding non-refreshed runs.
Ultimately, the refreshed scenario produces even higher charge separation $S\sim 100\%$, with nearly all $e^-$ confined inside the magnetized filaments and $e^+$ outside, as shown in Figure \ref{Fig:refresh_9000}b.

\begin{figure}[t]
\centering    
\vspace{0.2cm}
\includegraphics[width=0.49\textwidth,clip=true,trim= 0 700 0 0]{"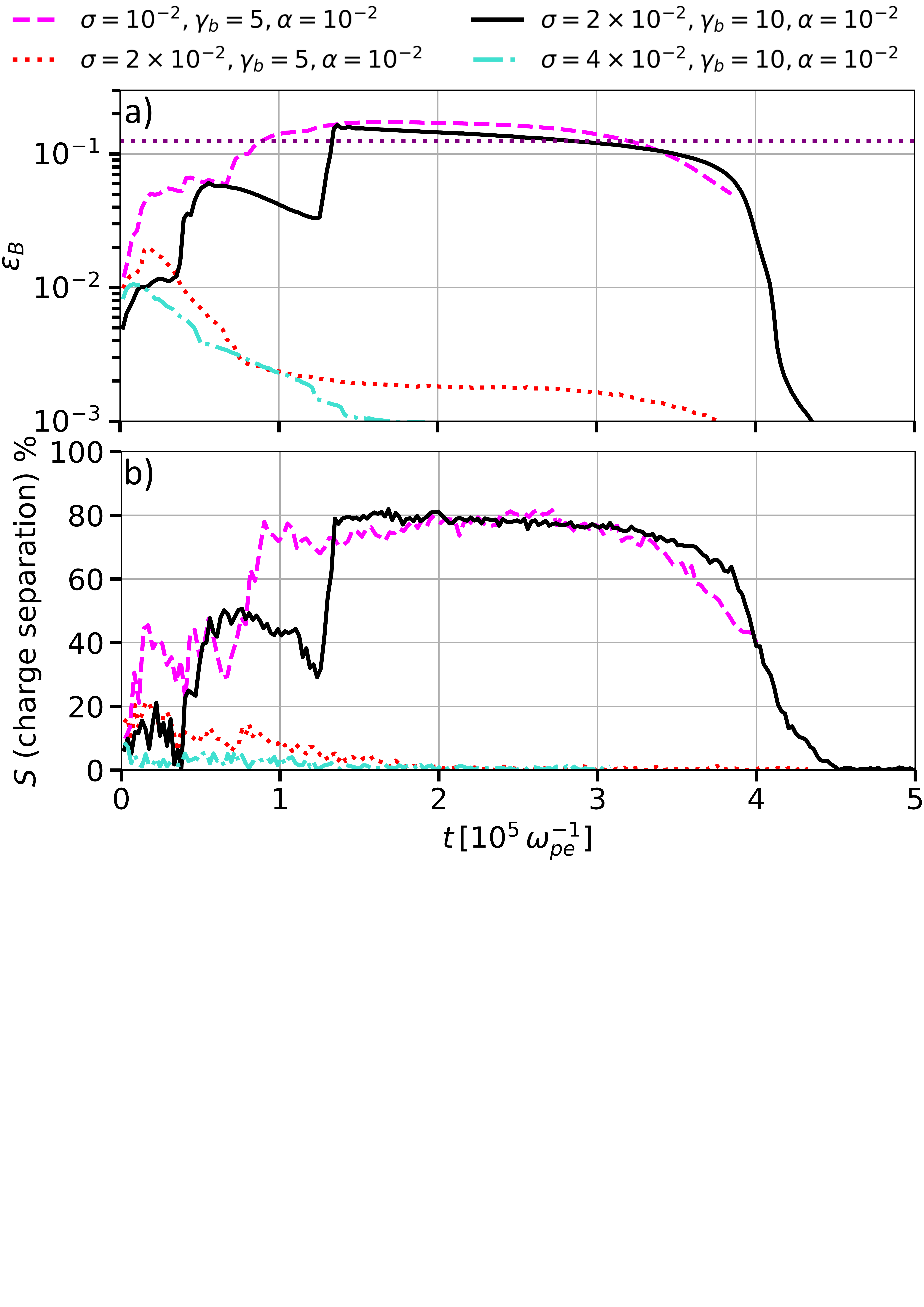"}
    \vspace{-58pt}
    \caption{Same as Figure \ref{Fig:no_refresh}, but for $(\alpha,\gamma_b)=(10^{-2},5)$ and $\sigma=10^{-2}$ (purple dashed), $(\alpha,\gamma_b)=(10^{-2},10)$ and $\sigma=2\times10^{-2}$ (solid black), $(\alpha,\gamma_b)=(10^{-2},5)$ and $\sigma=2\times10^{-2}$ (red dotted), and $(\alpha,\gamma_b)=(10^{-2},10)$ and $\sigma=4\times10^{-2}$ (turquoise dash-dotted).}
    \label{Fig:energetic}
\end{figure}

\smallskip
\noindent\textbf{\textit{Beam energy density vs. magnetic energy density.---}}
We confirm that our results depend on the ratio of the beam energy density to the magnetic energy density, rather than on the individual values of $\alpha$, $\sigma$, and $\gamma_b$. In the main text, we showed that two cases with the same $\alpha \gamma_b$ but different $\gamma_b$ and $\alpha$, specifically with ($\alpha ,\gamma_b$)$=(10^{-2},10$) and ($\alpha ,\gamma_b$)$=(5 \times 10^{-3},20$), at fixed $\sigma = 2 \times 10^{-2}$, lead to the development of the cavitation instability. Here, we further demonstrate that simultaneously decreasing $\sigma$ and $\alpha \gamma_b$ yields similar results in terms of $\epsilon_B$ and charge separation. In Fig.~\ref{Fig:energetic}, we show that for ($\alpha ,\gamma_b$)$=(10^{-2},5$), and $\sigma = 10^{-2}$, both $\epsilon_B$ and $S$ converge to values consistent with those obtained for ($\alpha ,\gamma_b$)$=(10^{-2},10$) and $\sigma=2 \times 10^{-2}$. Conversely, increasing $\sigma$ by a factor of two in both cases, thereby reducing $\alpha \gamma_b / \sigma$, suppresses the instability. This provides further support that the key parameter is $ \alpha \gamma_b/ \sigma$, and that our results can be rescaled to values of $\alpha$, $\gamma_b$ and $\sigma$ representative of the ISM, given that the value of $\alpha \gamma_b /\sigma$ in our simulations is similar to the one expected in X-ray filaments. In particular, Bell’s instability is expected to develop whenever this ratio satisfies the threshold condition for the onset of cavitation(see Eq. \ref{eq:bell}).

\smallskip
\noindent\textbf{\textit{Numerical convergence.---}}
We verified that our results are robust to changes in the number of particles per cell (ppc) and in the box size. This holds both for cases where the cavitation instability grows as well as where it does not grow; similarly, for both refreshed and non-refreshed cases.
In Fig.~\ref{Fig:numeric} we show $\epsilon_B$ and the degree of charge separation $S$  for our reference non-refreshed case. The four curves correspond to different box sizes, $48 \times 48 \, d_e^2$ (solid black) and $200 \times 200 \, d_e^2$ (dotted-dashed pink) and $1000 \times 1000 \, d_e^2$ (dotted gray), and to different ppc (dashed green). In all cases, the saturated magnetic field energy and the degree of charge separation converge to the same values. Differences appear only in the growth phase, where the larger boxes produces a smoother evolution due to the development of a larger number of cavities. The late-time decay also differs, being more gradual in the larger boxes, though the onset of decay is at roughly the same time.

\begin{figure}[t]
\centering    
\vspace{0.2cm}
\includegraphics[width=0.49\textwidth,clip=true,trim= 100 10 0 0]{"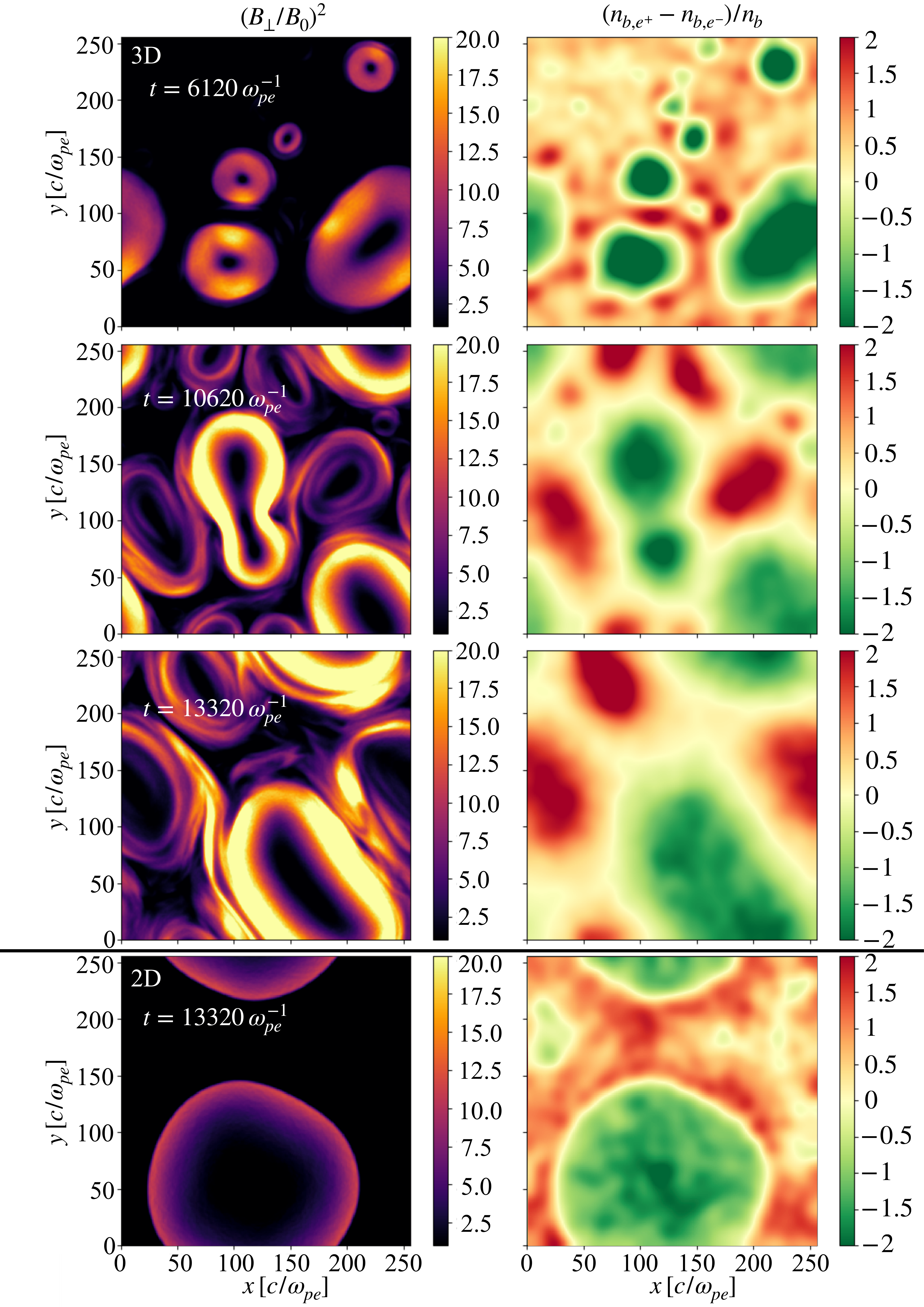"}
    \caption{Evolution of a slice in the $x$–$y$ plane showing $(B_\perp/B_0)^2$ (first column) and of the charge separation (second column) of beam positrons and electrons $(n_{b,e^+}-n_{b,e^-})/n_b$ for the 3D simulation with $\alpha \gamma_b = 5 \times 10^{-1}$, $\sigma = 10^{-1}$, and $\Delta \gamma_b = 5$, compared with an equivalent 2D simulation (bottom row).} 
    \label{Fig:3d_supplemental}
\end{figure}

\smallskip
\noindent\textbf{\textit{3D simulation.---}}
As discussed in the main text, we perform a 3D simulation with $\gamma_b=10$, $\alpha=5\times 10^{-2}$, $\sigma=10^{-1}$, and $\Delta \gamma_b=5$, motivated by \cite{zacharegkas+24}, which predicts that hot beams drive the NRI more efficiently than cold ones. In Fig.~\ref{Fig:3d_supplemental} we show the evolution of a slice in the $x$–$y$ plane of $(B_\perp/B_0)^2$ (first column) and the charge separation (second column) of beam positrons and electrons $(n_{b,e^+}-n_{b,e^-})/n_b$ for the 3D case. The bottom row shows, at the final time, the results of a 2D simulation with the same physical and numerical parameters as the 3D run.
The cavitation instability initially grows, forming toroidal structures around the beam. At later times, $e^+$ outside the cavities further enhance the magnetic field via the NRI, generating right-hand circularly polarized waves with wavevector along $z$. 
This secondary amplification occurs only in 3D; in 2D, the $e^+$-driven modes do not develop (bottom row of Fig.~\ref{Fig:3d_supplemental}), consistent with the NRI requiring wavevectors along the beam direction.

\end{document}